\documentclass[preprint,5p]{elsarticle}
\usepackage{hyperref}
\usepackage{natbib}
\usepackage{graphicx}
\usepackage{algpseudocode}
\makeatletter
\renewcommand{\ALG@beginalgorithmic}{\small}
\makeatother
\usepackage{algorithm}
\usepackage[strings]{underscore}
\usepackage{xurl}
\usepackage{amsmath}
\usepackage{amssymb}
\usepackage{array}
\usepackage{multirow}
\usepackage{subfigure}
\usepackage{float}
\usepackage{caption}
\biboptions{sort&compress}

\journal{Computer Networks}

\begin{document}

\begin{frontmatter}

\title{Generative Adversarial Networks (GANs) in Networking: A Comprehensive Survey \& Evaluation}

\author[1]{Hojjat~Navidan}
\address[1]{School of Electrical and Computer Engineering, College of Engineering, University of Tehran, Tehran 14395-515, Iran}

\author[2]{Parisa~Fard~Moshiri}
\author[2]{Mohammad~Nabati}
\address[2]{Cognitive Telecommunication Research Group, Department of Telecommunications, Faculty of Electrical Engineering, Shahid Beheshti University, Tehran 19839-69411, Iran}

\author[3] {Reza~Shahbazian}
\address[3] {Department of Electrical Engineering, Faculty of Technology and Engineering, Standard Research Institute, Alborz 31745-139, Iran}

\author [2,4] {Seyed~Ali~Ghorashi\corref{cor1}\fnmark[fn1]}
\address[4]{School of Architecture, Computing and Engineering, University of East London, E16 2RD London, U.K}

\author [1] {Vahid~Shah-Mansouri}
\author [5] {David~Windridge}
\address [5] {Department of Computer Science, School of Science and Technology, Middlesex University, NW4 4BT London, U.K}

\cortext[cor1]{Corresponding author}
\fntext[fn1]{email: s.a.ghorashi@uel.ac.uk}

\begin{abstract}
Despite the recency of their conception, Generative Adversarial Networks (GANs) constitute an extensively-researched machine learning sub-field for the creation of synthetic data through deep generative modeling. GANs have consequently been applied in a number of domains, most notably computer vision, in which they are typically used to generate or transform synthetic images. Given their relative ease of use, it is therefore natural that researchers in the field of networking (which has seen extensive application of deep learning methods) should take an interest in GAN-based approaches. The need for a comprehensive survey of such activity is therefore urgent. In this paper, we demonstrate how this branch of machine learning can benefit multiple aspects of computer and communication networks, including mobile networks, network analysis, internet of things, physical layer, and cybersecurity. In doing so, we shall provide a novel evaluation framework for comparing the performance of different models in non-image applications, applying this to a number of reference network datasets.
\end{abstract}

\begin{keyword}Generative Adversarial Networks\sep Deep Learning\sep Semi-supervised Learning\sep Computer Networks\sep Communication Networks.\end{keyword}
\end{frontmatter}

\section{Introduction}
\label{sec:1}
Over the past few years, there has been an exponential growth of mobile networks. According to the Ericsson Mobility Report, there are more than 7 billion mobile broadband subscriptions extant in the world. With the rapid uptake of the fifth-generation (5G) network, it is estimated that by the end of 2025, this number will increase to 9 billion \cite{EricssonMobilityReport}. This growth has additionally led to an enormous rise in network and infrastructure demands. For example, 5G systems are designed to support massive traffic volumes, real-time network analysis, and agile management of resources; however, since mobile networks are heterogeneous, complex, and non-linear, meeting these requirements with classic methods and algorithms becomes challenging. Many of the problems encountered in computer and communication networks can be formulated as classification, detection, estimation, prediction, or optimization problems. Moreover, with advancements in processing and computing power, machine learning algorithms have increasingly demonstrated the ability to solve such problems more effectively than alternative approaches. Researchers are consequently led to propose new algorithms and methods based on machine learning and related data-driven approaches on a regular basis in order to overcome these challenges \cite{zhangDeepLearningMobile, jagannathMachineLearningWireless}.

\begin{table}[t]
	\centering
	\renewcommand{\arraystretch}{1.5}
	\caption{Full list of abbreviations in alphabetical order.}
	\label{tab:1}
	\resizebox{\columnwidth}{!}{%
		\begin{tabular}{|c|c|}
			\hline
			\textbf{Abbreviation} & \textbf{Explanation}                                      \\ \hline
			5G            & Fifth-generation mobile network                                    \\ \hline
			ACGAN         & Auxiliary   Classifier Generative Adversarial Network              \\ \hline
			API           & Application Programming Interfaces                                 \\ \hline
			AWGN          & Additive   White Gaussian Channel                                  \\ \hline
			BIGAN         & Bidirectional Generative Adversarial Network                       \\ \hline
			CGAN          & Conditional Generative Adversarial Network                         \\ \hline
			CNN           & Convolutional Neural Network                                       \\ \hline
			CSI           & Channel State Information                                          \\ \hline
			EMD           & Earth Mover Distance                                               \\ \hline
			GAN           & Generative   Adversarial Network                                   \\ \hline
			GCN           & Graph   Convolutional Network                                      \\ \hline
			GIDS          & GAN-based Intrusion Detection System                               \\ \hline
			HAR           & Human Activity Recognition                                         \\ \hline
			IDS           & Intrusion Detection System                                         \\ \hline
			IoT           & Internet of Things                                                 \\ \hline
			KDE           & Kernel Destiny Estimation                                          \\ \hline
			LSGAN         & Least Square Generative Adversarial Network                        \\ \hline
			LSTM          & Long   Short-Term Memory                                           \\ \hline
			MLP           & Multilayer   Perception                                            \\ \hline
			MMD           & Maximum Mean Discrepancy                                           \\ \hline
			NN           & Neural Network                                                      \\ \hline
			PDF           & Probability Density Function                                       \\ \hline
			QOE           & Quality of Experience                                              \\ \hline
			RCGAN         & Radio   Classify GAN                                               \\ \hline
			RF            & Radio Frequency                                                    \\ \hline
			RSS           & Received Signal Strength                                           \\ \hline
			RSSI          & Received Signal Strength Indicator                                 \\ \hline
			SAE           & Sparse Autoencoder                                                 \\ \hline
			SAGA          & Spectrum   Augmentation/Adaptation with GAN                        \\ \hline
			SON           & Self-Organizing Network                                            \\ \hline
			UAV           & Unnamed   Aerial Vehicles                                          \\ \hline
			WGAN          & Wasserstein Generative Adversarial Network                         \\ \hline
			WGAN-GP       & Wasserstein Generative Adversarial Network with   Gradient Penalty \\ \hline
			WSN           & Wireless Sensor Networks                                           \\ \hline
		\end{tabular}%
	}
\end{table}

\par These learning algorithms, however, often require a considerable amount of training data to be effective. In many real-world problems, data accessibility is limited, or it may be prohibitively expensive to gather sufficient amounts of data. A further issue is that many real-world data scenarios involve significant skew in the underlying class distributions, making obtaining representative data more complex; as many machine learning algorithms are also adversely affected by class imbalance \cite{krawczykLearningImbalancedData}. 
\par Semi-supervised learning is a branch of machine learning that attempts to address the problem of partially labeled training data. Data labeling is frequently expensive and time-consuming, and label storage may present problems in some cases. Problems that involve diverse, unstructured and inter-connected datasets fall into this category \cite{zhangSurveyDeepLearning}. Grappling with the problems of data shortage, class imbalance, and label propagation has naturally prompted consideration of generative approaches; that is, using innovative methods for generating the new data with the same properties as real-data in order to improve the performance of learning algorithms. 
\par Deep generative models have emerged as one of the most exciting and prominent sub-fields of deep learning, given their remarkable ability to synthesize input data of arbitrary form by learning the distribution such that novel samples can be drawn. Deep generative models
can consequently provide a variety of benefits. Firstly, supervised learning methods often require a substantial quantity of data in order to achieve good performance (in many real-case scenarios such as indoor localization, the requisite amount of data may not be accessible \cite{nabatiUsingSyntheticData}); secondly, collection of large quantities of data may be infeasibly time-consuming or expensive. Besides alleviating difficulties in these single-domain scenarios, deep generative approaches have also demonstrated their usefulness in transfer learning scenarios, where the correlation between two datasets is utilized in conjunction with generative models to transfer learning between datasets \cite{zhuMultimodalImagetoImageTranslation}.
\par From the first introduction of Generative Adversarial Networks in 2014, GANs have been a focus of attention in generative machine learning (according to Google scholar, there are around 75000 papers based or focusing on GANs to date\footnote{November 2020}). GANs have predominantly been used in computer vision, including but not limited to image generation, face synthesis \cite{debAdvFacesAdversarialFace}, image translation \cite{isolaImagetoImageTranslationConditional, zhuUnpairedImagetoImageTranslation, yiDualGANUnsupervisedDual}, texture synthesis \cite{fadaeddiniCaseStudyGenerative, jetchevTextureSynthesisSpatial}, medical imaging, \cite{yiGenerativeAdversarialNetwork} and super-resolution \cite{ledigPhotoRealisticSingleImage}. Moreover, GANs can be applied in many other fields including but not limited to voice and speech signals \cite{liSpeechBandwidthExtension, chenGeneratingMusicAlgorithm, daiEndtoendGenerativeNetwork}, anomaly detection \cite{dimattiaSurveyGANsAnomaly}, power systems and smart grids \cite{yingPowerMessageGeneration, bagheriGenerativeAdversarialModelGuided, zhangGenerativeAdversarialNetwork}, electronics \cite{wangAdaBalGANImprovedGenerative, xuWellGANGenerativeAdversarialNetworkGuidedWell}, and fault diagnosis \cite{zhouDeepLearningFault, wangGeneralizationDeepNeural, xieTransferLearningStrategy, zhaoAeroEngineFaultsDiagnosis}.
\par In line with this activity across multiple research fronts, there has been extensive recent research interest in applying generative adversarial networks to computer and communication networks; hence the need for a comprehensive survey covering the full extent of this new field of development. 
\subsection{Previous Work}
A number of surveys and tutorials on GANs within the broader field of machine learning exist. For example, Cao et al. \cite{caoRecentAdvancesGenerative} introduced the most frequently-used GAN models and compared functionality with respect to a range of use cases. They conducted experiments to generate synthetic images from the two widely used image datasets, MNIST \cite{lecunGradientbasedLearningApplied} and Fashion-MNIST \cite{xiaoFashionMNISTNovelImage}, evaluating the performance of the different GANs models both visually and quantitatively. Some of the other surveys have touched upon fields outside of computer vision; for example, Wang et al. \cite{wangGenerativeAdversarialNetworks} briefly reviewed applications of GANs to other areas of interest, including speech and language processing.
\par Table \ref{tab:2} summarizes the relationship between our survey and the other extant surveys and tutorials. As indicated, while most current survey works covering GANs extend across the field of computer vision, a few cover other fields. Our survey will thus seek to position itself in relation to these extant surveys by providing a detailed overview of adversarial networks' applicable to the field of computer and communication networks. In doing so, we shall introduce several novel evaluation metrics that can be deployed in relation to the generation of non-image data in order to evaluate the performance of different GAN models.

\begin{table*}
	\centering
	\renewcommand{\arraystretch}{1.3}
	\caption{Comparison of current survey with existing surveys and tutorials. ``CV'' and ``NET'' refer to computer vision and networking.}
	\label{tab:2}
	\resizebox{\textwidth}{!}{%
		\begin{tabular}{|c|c|c|c|c|c|c|c|} 
			\hline
			\multirow{2}{*}{Existing Publications} & \multicolumn{4}{c|}{Technical Overview}               & \multicolumn{3}{c|}{Applications Overview}  \\ 
			\cline{2-8}
			& Models Intro. & Models Comp. & Eval. Metrics & Simulation & CV & NET & Others                           \\ 
			\hline
			Pan et al. \cite{panRecentProgressGenerative}                        & \checkmark             & \checkmark            & \checkmark             &        & \checkmark       &          &             \\ \hline
			Turhan and Bilge \cite{turhanRecentTrendsDeep}                  & \checkmark             & \checkmark            &               &        &         &          &             \\ \hline
			Cao et al. \cite{caoRecentAdvancesGenerative}                        & \checkmark            & \checkmark           & \checkmark    & \checkmark      & \checkmark       &         &             \\ \hline
			Goodfellow \cite{goodfellowNIPS2016Tutorial}                       & \checkmark            &              &               &        & \checkmark       &          &             \\ \hline
			Gonog and Zhou \cite{gonogReviewGenerativeAdversarial}                    & \checkmark             & \checkmark           &               &        & \checkmark      &         & \checkmark          \\ \hline
			Zhang et al. \cite{zhangRecentAdvanceGenerative}                     & \checkmark            & \checkmark            &               &        &         &          &             \\ \hline
			Wu et al. \cite{wuSurveyImageSynthesis}                        & \checkmark            &              &               &        & \checkmark      &          &             \\ \hline
			Wang et al. \cite{wangGenerativeAdversarialNetworks}                      & \checkmark             & \checkmark            &               &        & \checkmark       &         & \checkmark           \\ \hline
			Shorten and Khoshgoftaar \cite{shortenSurveyImageData}         & \checkmark            & \checkmark           &               &        & \checkmark      &          &             \\ \hline
			Esfahani and Latifi \cite{nasresfahaniImageGenerationGansbased}              & \checkmark             &              &               &        & \checkmark       &          &             \\ \hline
			Creswell et al. \cite{creswellGenerativeAdversarialNetworks}                  & \checkmark             & \checkmark            &               &        & \checkmark       &         &             \\ \hline
			Di Mattia et al. \cite{dimattiaSurveyGANsAnomaly}                 & \checkmark             & \checkmark            &               & \checkmark      &         &         &\checkmark           \\ \hline
			Our Survey                             & \checkmark             & \checkmark           & \checkmark            & \checkmark     &         &\checkmark        & \checkmark           \\ \hline
		\end{tabular}%
	}
\end{table*}

\subsection{Key Contributions}
To the best of our knowledge, there is no survey or tutorial specifically discussing recent developments of GANs in relation to computer and communication networks. As indicated above, the majority of the research done in this area covers computer vision and image processing. This fact motivates us to provide a comprehensive survey and review of recent relevant researches carried out in the field of networking. We additionally provide an evaluation framework to measure and compare the performance of the respective GAN models. The key contributions of this survey paper are summarized as follows:
\begin{itemize}
	\item We provide a comprehensive but compact background concerning deep generative models, with an emphasis on GANs. While different GAN variants have a similar underlying mechanism, network architectures significantly differ. We therefore compare and contrast network architectural components in detail.
	\item We discuss the range of applications that have benefited from GANs in the literature categorized into five main categories, depending on whether the GANs in question produce synthetic data for semi-supervised learning or else utilize the generator or the discriminator network in a unique way. For each such category, we describe in detail how GANs seek to provide a solution for some of the existing challenges.
	\item Inspired by evaluation metrics for image data, we provide a framework for comparing the performance of differing GAN models trained on four network datasets of different types. Furthermore, we visualize the data distributions and utilize statistical tools to compare the similarity between them. To the best of our knowledge, this is the first time that the performance of GANs has been evaluated comparatively for non-image data (since GANs have predominantly been used in computer vision, principally for the generation of images).
	\item We discuss the open challenges of GANs and indicate how addressing these challenges has the potential to contribute to further improvements in networking. We end by suggesting some research directions and areas with considerable potential to benefit from a GAN-based approach but which have not yet exploited their potential.
\end{itemize}

\subsection{Survey Organization}
The remainder of this paper is organized as follows: We begin by providing a background concerning deep generative methods in section \ref{sec:2}, in particular Generative Adversarial Networks (GANs). We provide details of a number of the main state-of-the-art GAN networks for non-image data. Next in section \ref{sec:3}, we review recent GAN applications in computer and communication networks, grouped under mobile networks, network analysis, internet of things, physical layer, and cybersecurity. In section \ref{sec:4}, we introduce a framework for evaluating the performance of different GAN variants and conduct experiments to compare various state-of-the-art models such as CGAN, LSGAN, INFOGAN, and WGAN. Finally, we discuss current challenges and future work to conclude the paper.

\section{Deep Generative Models}
\label{sec:2}
A generative model is defined as any model that can represent an estimation of the given data probability distribution by drawing samples from it. These models either result in a distribution that estimates the original model explicitly or else generates samples from the original data without defining a distribution \cite{goodfellowNIPS2016Tutorial}. They have a wide area of application, including reinforcement learning \cite{gamrianTransferLearningRelated} and inverse reinforcement learning \cite{finnConnectionGenerativeAdversarial} in which they are used to simulate possible scenarios for multi-modal learning \cite{pandevaMMGANGenerativeAdversarial}. However, the predominant application of these models is the generative filling-in of missing data and data imputation. They can hence be utilized in many scenarios, such as semi-supervised learning, in which only a portion of training data is labeled. As most modern deep learning models and algorithms require extensive labeled examples for training, semi-supervised learning provides a ready solution for reducing the labeling requirement. GANs, in particular, have found extensive use in semi-supervised learning \cite{kumarSemisupervisedLearningGANs}.
\par
\begin{figure}
	\centering
	\includegraphics[width=\linewidth]{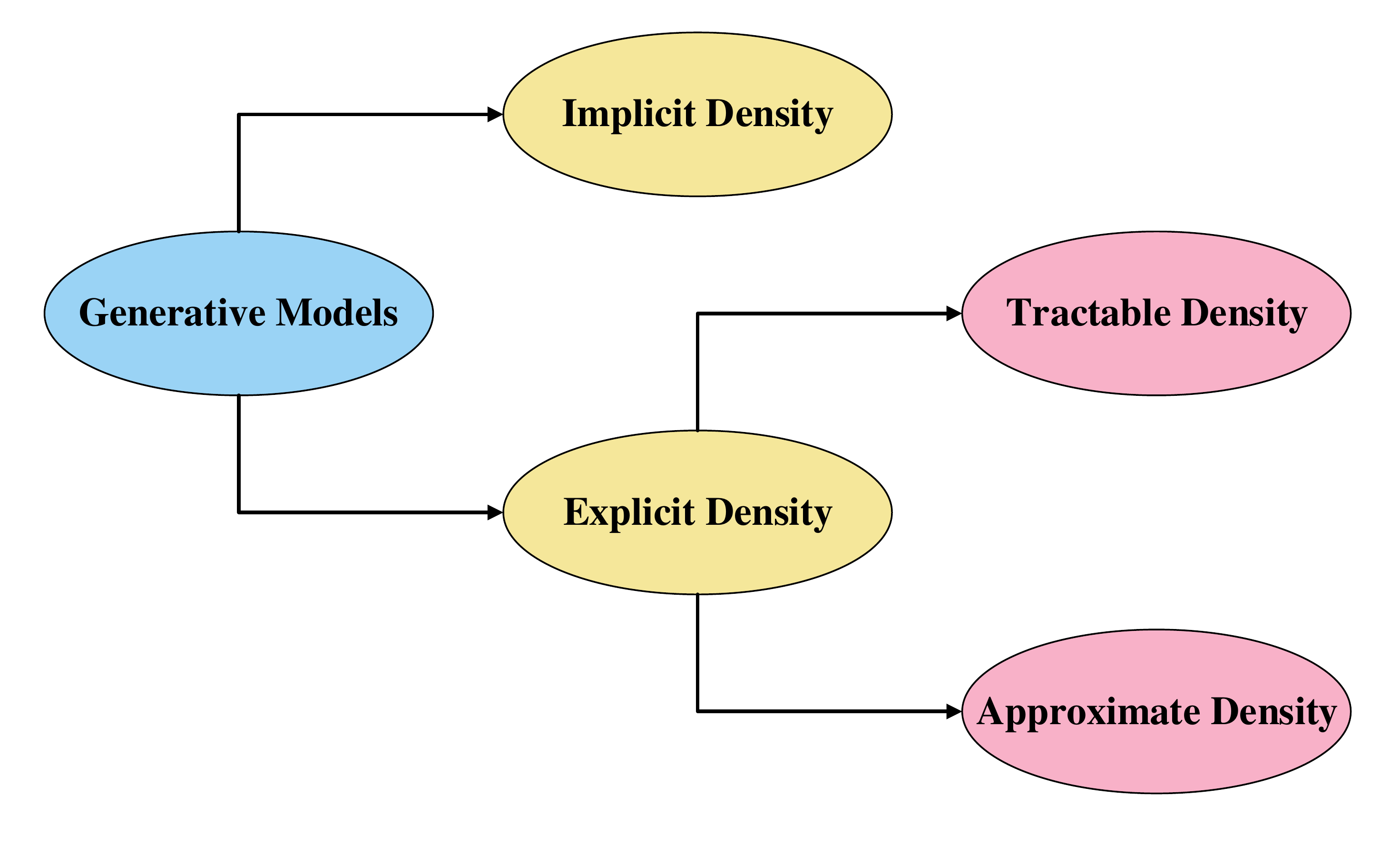}
	\caption{Overall taxonomy of generative models.}
	\label{fig:1}
\end{figure}

 The overall taxonomy of generative models is depicted in Fig.\ref{fig:1}. Broadly, generative models can be divided into two main types: explicit density and implicit density models. Explicit density models are those that provide an explicit parametric specification of the data distribution. The main challenge here is capturing all the complexity of the data while maintaining computational tractability. Consequently, explicit density models are, in turn, divided into two sub-groups. Firstly, there are the models that define computationally-tractable density functions, such as deep belief networks \cite{keyvanradBriefSurveyDeep} or flow models \cite{rezendeVariationalInferenceNormalizing}. These models allow us to use an optimization algorithm directly on the log-likelihood of training data and hence be highly effective; however, they also have intrinsic limitations, resulting in a range of practical drawbacks depending on the data distribution. Secondly, there are the models with intractable density functions that use approximations, either variational (e.g., variational autoencoders) or Monte Carlo based (e.g., Boltzmann machines), to maximize the likelihood. In contrast to these two explicit subclasses, implicit density models do not specify the distribution of data and thus do not require a tractable likelihood but rather seek to define a stochastic process that aims to draw samples from the target data distribution. GANs are implicit density generative models of this latter kind which are able to generate data samples in a single step \cite{goodfellowNIPS2016Tutorial}. 

\subsection{Generative Adversarial Networks}
In 2014, Goodfellow et al. \cite{goodfellowGenerativeAdversarialNets} introduced a novel adversarial class of generative models, GANs, which aim to produce synthetic data with maximal similarity to the original data. The GAN model consists of two main aspects, a Generator and a Discriminator, the idea behind this model being intrinsically game-theoretic, albeit within a deep learning context. The generator hence has the role of a counterfeiter, aiming to deceive the discriminator. Countering this, the discriminator plays a policing role that aims to recognize the counterfeits. Consequently, both the generator and discriminator learn from each other in developing their capabilities. After termination of the learning stage, the generator is a fully-trained counterfeiter able to maximally mislead the `police,' while the discriminator is maximally trained to realize counterfeits. Some of the key methodological variants of GANs will be covered in the next subsection.

\subsection{GAN Variations and Architectures}
\begin{figure*}[!h]
	\centering
	\includegraphics[width=\linewidth]{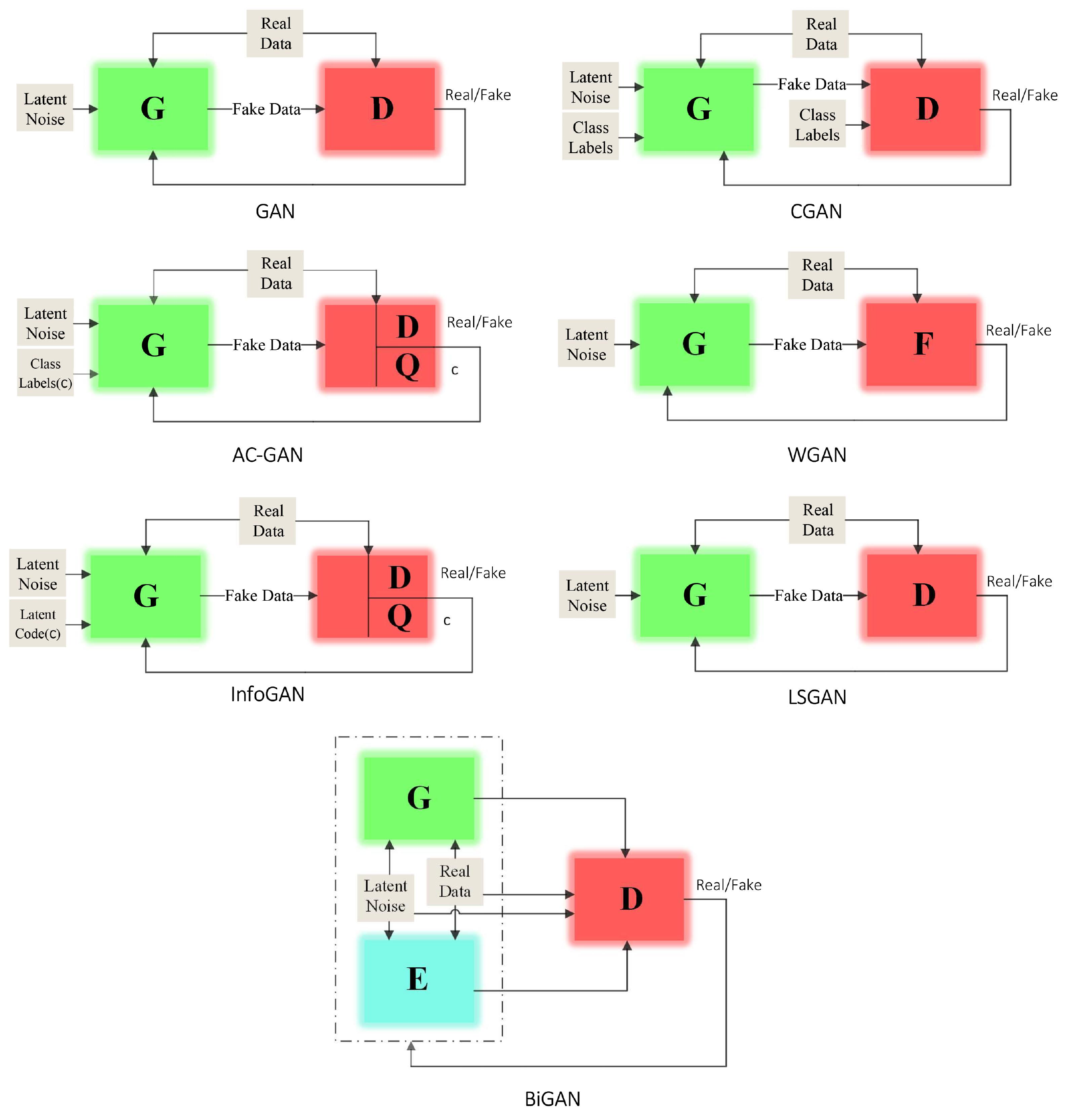}
	\caption{Architecture of the GANs reviewed in this section. G refers to the generator network, which is present in all of the above models. D refers to the discriminator network while E, F, and Q refer to the encoder, critic, and the Q-network respectively.}
	\label{fig:2}
\end{figure*}

The model initially proposed by Goodfellow (from now on referred to as Vanilla GAN) suffers from a few significant problems, such as vanishing gradients, mode collapse, and a failure to converge \cite{salimansImprovedTechniquesTraining}. Many solutions have since been proposed to overcome these challenges: some focus on improving the training method to prevent these problems from occurring, while others constitute entirely different architectures. Examples of the latter include Conditional GAN (CGAN) \cite{mirzaConditionalGenerativeAdversarial}, CycleGAN \cite{zhuUnpairedImagetoImageTranslation}, Bidirectional GAN (BiGAN) \cite{donahueAdversarialFeatureLearning}, DualGAN \cite{yiDualGANUnsupervisedDual}, DiscoGAN \cite{kimLearningDiscoverCrossDomain}, Pix2Pix \cite{isolaImagetoImageTranslationConditional}, InfoGAN \cite{chenInfoGANInterpretableRepresentation}, Energy-based GAN (EBGAN) \cite{zhaoEnergybasedGenerativeAdversarial}, Wasserstein GAN (WGAN) \cite{arjovskyWassersteinGAN}, and Super-Resolution GAN \cite{ledigPhotoRealisticSingleImage}. Of these proposed models, only a few have been used for data other than images; including but not limited to Vanilla GAN, BIGAN, CGAN, InfoGAN, CycleGAN, EBGAN, and Least Square GAN (LSGAN) \cite{maoLeastSquaresGenerative}. The other models are either exclusively designed for image data or else have not been used in applications other than computer vision. For example, Deep Convolutional GAN (DCGAN) is a robust variant of GANs that utilizes Convolutional Neural Networks (CNNs) to generate high-quality images \cite{radford2016unsupervised}; however, as they contain a CNN, they are only applicable where the sequential spatial features are
of importance (as is the case for image data). Below, we introduce seven classes of general-use GAN models described in the literature, focusing particularly on those especially relevant to the network domain. The architectures of these models are depicted in Fig.\ref{fig:2}.
\par We shall first describe the learning process of the Vanilla GAN in Fig.\ref{fig:2}. As may be seen, the architecture of this model consists of two components, the G and D networks. Arbitrary latent noise spikes in the G network are used to generate fake samples. Afterwards, the D network compares the fake and real samples to increase its own discrimination performance. We can implement both $G$ and $D$ (which are differentiable functions) via a Multilayer Perception (MLP). The process of learning these two networks is hence based on the minimax cost function, which is defined as follows:
\begin{equation}
\label{eq:1}
\begin{split}
\mathop {\min }\limits_G \mathop {\max }\limits_D \,\,L(D,G)  =& {E_{x\sim{p_{data}}(x)}}[\log D(\mathbf{x})]  \\  +&{E_{z\sim{p_z}(z)}}[\log (1 - D(G(\mathbf{z}))],
\end{split}
\end{equation}
where ${p_{data}(x)}$ is the probability distribution of the data and ${p_z}(z)$ is the prior noise distribution. 
\par This cost function, which aims to maximize the probability of correctly assigning labels to real and fake data, consists of two parts. The generator thus seeks to map the latent noise $\mathbf{z}\sim {p_z}(z)$ to the real data distribution, which is denoted via $G(\mathbf{z;}{\theta _g})$, where ${\theta _g}$ indicates the parameters of the generator.
Simultaneously, the discriminator is trained to distinguish between real and fake data, denoted via $D(\mathbf{x;}{\theta _d})$, where ${\theta _d}$ indicates the parameters of the discriminator. Naturally, the discriminator's output is binary, where 0 and 1 specify fake and real data, respectively. By fixing the respective sides relating to each model, the opposing model's cost function can be represented. Hence, the discriminator's loss is defined as follows:
\begin{equation}
\label{eq:2}
\begin{split}
L({\theta _d}) =& {E_{x\sim{p_{data}}(x)}}[\log D(\mathbf{x;}{\theta _d})] \\ +& {E_{z\sim{p_z}(z)}}[\log (1 - D(G(\mathbf{z;}{\theta _g}))].
\end{split}
\end{equation}
\par It should be noted that since the first term of Eq. (\ref{eq:1}) does not affect the generator when it is fixed, it disappears in the gradient updating step. Similarly, the loss function of the generator is defined as follows:
\begin{equation}
\label{eq:3}
L({\theta _g}) = {E_{z\sim{p_z}(z)}}[\log (1 - D(G(\mathbf{z;}{\theta _g}))].
\end{equation}
\par The process of updating ${\theta _d}$ and ${\theta _g}$ is depicted in Algorithm 1 \cite{nabatiUsingSyntheticData}. Convergence occurs when $D(\mathbf{x;}{\theta _d}) = 0.5$, which means that discriminator is no longer able to distinguish between real and fake data. Once the algorithm has converged, the generator can then produce synthetic data by random sampling of the same prior noise distribution $\mathbf{z}\sim{p_z}(z)$, such that the discriminator is not able to distinguish whether these data are real or not. This model is hence the baseline structure of the vast majority of GANs model variants that researchers have introduced in recent years.

\begin{algorithm}[t!]
	\label{alg:1}
	\caption{GAN learning process}
	\begin{algorithmic}[1]
		\State Initiate {${\begin{array}{*{20}{l}}
				{{e  =  \textrm{epochs},}\,\,{s  =  \textrm{step size for updating }}{\rm{ }}{\theta _d}},\\
				{\textrm{learning rates (}\,{}{\rm{ }}{\eta _g}\,\textrm{and }{\eta _d}{)}}
				\end{array}}$}
		\For {$i= 1:e$}
		\For {{$t = 1:s$}}
		\State {{sample batch } ${\mathbf{Z}} \in {{\bf{R}}^{B \times L}} \sim {\mathbf{p}_z}\mathbf{(z)}$}
		\State {sample batch $\mathbf{X} \in {{\bf{R}}^{B \times M}}$ from data}
		\State {update $\theta_d $ by gradient ascent based optimizer}
		\State {$L({\theta _d}) = \frac{1}{B}\sum\limits_{\mathbf{b} = 1}^\mathbf{B} {\left[ {\log {D}({\mathbf{X}_b}\mathbf{,}{\theta _d}) + \log (1 - {D}(\mathbf{G}({\mathbf{Z}_b}\mathbf{,}{\theta _g}))} \right]} $}
		\State {${\psi _d} = \frac{\partial }{{\partial {\theta _d}}}L({\theta _d})$}
		\State {$\theta _d^{{\rm{t}} + 1} = \theta _d^{\rm{t}} + \mathbf{ }{\eta _d}{\psi _d}$}
		\EndFor 
		\State {{sample batch } ${\mathbf{Z}} \in {{\bf{R}}^{B \times L}} \sim {\mathbf{p}_z}\mathbf{(z)}$}
		\State {update $\theta_g $ by gradient descent based optimizer}
		\State {$\,\,\,\,\,L({\theta _g}) = \frac{1}{B}\sum\limits_{b = 1}^B {\log (1 - D(G({\mathbf{Z}_b}\mathbf{,}{\theta _g}))} $}
		\State {${\psi _g} = \frac{\partial }{{\partial {\theta _g}}}L({\theta _g})$}
		\State {$\theta _g^{{\rm{t}} + 1} = \theta _g^{\rm{t}} - \mathbf{ }{\eta _g}{\psi _g}$}
		\EndFor
		\State {${\mathbf{X}_b}$\,and\,${\mathbf{Z}_b}$\,\,are\,\,${b}'$th\,\,rows\,\,of\,\,$\mathbf{X}$\,\,and\,\,$\mathbf{Z}$,\,\,respectively.}
		
	\end{algorithmic}
\end{algorithm}

\subsubsection{Conditional GAN}
The Vanilla GAN can only generate data via the given inputs and cannot generate samples with labels simultaneously. Therefore, Mirza and Osindero \cite{mirzaConditionalGenerativeAdversarial} proposed Conditional GAN (CGAN), which feeds relevant additional information to the generator and discriminator sides by presenting the encoded class labels alongside the prior noise and real data, respectively. The cost function, very similar to the Vanilla GAN, is defined as follows:
\begin{equation}
\label{eq:4}
\begin{split}
\mathop {\min }\limits_G \mathop {\max }\limits_D \,\,L(D,G)  =& {E_{x\sim{p_{data}}(x)}}[\log D(\mathbf{x}|\mathbf{y})]  \\ +&{E_{z\sim{p_z}(z)}}[\log (1 - D(G(\mathbf{z}|\mathbf{y}))].
\end{split}
\end{equation}
Consequently, the discriminator's loss can be formulated as:
\begin{equation}
\label{eq:5}
\begin{split}
L({\theta _d}) =& {E_{x\sim{p_{data}}(x)}}[\log D(\mathbf{x}|\mathbf{y;}{\theta _d})] \\ +& {E_{z\sim{p_z}(z)}}[\log (1 - D(G(\mathbf{z}|\mathbf{y;}{\theta _g}))],
\end{split}
\end{equation}
and the generator's loss is defined as:
\begin{equation}
\label{eq:6}
\begin{split}
L({\theta _g}) = {E_{z\sim{p_z}(z)}}[\log (1 - D(G(\mathbf{z}|\mathbf{y;}{\theta _g}))].
\end{split}
\end{equation}

\subsubsection{Auxiliary Classifier GAN}
Odena et al. \cite{odenaConditionalImageSynthesis} propose a variant of the GAN architecture named Auxiliary Classifier GAN (ACGAN). In ACGAN, the generated samples have a corresponding class label $\mathbf{c}\sim{p_c}(c)$ alongside the noise parametrization $\mathbf{z}\sim{p_z}(z)$. The generator uses both of these to generate synthetic samples. Besides being responsible for distinguishing real and fake data, the discriminator must also carry out a task $(Q)$, predicting the class labels. The significant difference between ACGAN and CGAN is that CGAN has conditioning on class labels at the input of the generator and the discriminator to generate data for each class. ACGAN, however, predicts class labels via a multi-task architecture consisting of a paired source and label loss as follows:
\begin{equation}
\label{eq:7}
\begin{split}
&{L_S} = {E_{x\sim{p_{data}}(x)}}[\log D(\mathbf{x})] + {E_{z\sim{p_z}(z)}}[\log (1 - D(G(\mathbf{z}))]\\
&{L_C} = {E_{x\sim{p_{data}}(x)}}[\log Q(\mathbf{c}|\mathbf{x})] + {E_{z\sim{p_z}(z)}}[\log (Q(\mathbf{c}|\mathbf{z})].
\end{split}
\end{equation}
\par The generator hence aims to maximize ${L_S} + {L_C}$, while on the other hand the discriminator seeks to maximize ${L_S} - {L_C}$.
\subsubsection{Wasserstein GAN}
The Vanilla GAN suffers from vanishing gradient and convergence problems making the training stage inconvenient. Various solutions have been proposed in recent years to overcome these issues. Arjovsky et al. \cite{arjovskyPrincipledMethodsTraining} suggest adding additional noise to the generated samples to better stabilize the model before presentation to the discriminator. The same group also proposed a novel cost function \cite{arjovskyWassersteinGAN} to deal with instability problems, the resulting model being the Wasserstein GAN (WGAN) in which the cost function is given as:
\begin{equation}
\label{eq:8}
\begin{split}
L(F,G) = \mathop {\sup }\limits_{||F|{|_L} \le 1} {E_{x\sim{p_{data}}(x)}}[F(\mathbf{x})] - {E_{z\sim{p_z}(z)}}[F(G(\mathbf{z}))],
\end{split}
\end{equation}
where “sup” is the supremum over all the 1-Lipschitz functions with the constraint $\left| {F({x_1}) - F({x_2})} \right| \le \left| {{x_1} - {x_2}} \right|$  \cite{sohrabBasicRealAnalysis}.
\subsubsection{WGAN-GP}
Wasserstein GAN improves convergence and presents a solution to instability problems in the Vanilla GAN. However, it suffers from the undesirable behavior of critic weight clipping in the training stage. Gulrajani et al. \cite{gulrajaniImprovedTrainingWasserstein} add a gradient penalty term to the cost function of WGAN to overcome this bottleneck, the resulting model being referred to as `WGAN with Gradient Penalty' (WGAN-GP), which has the cost function:
\begin{equation}
\label{eq:9}
\begin{split}
L(F,G) =& \mathop {\sup }\limits_{||F|{|_L} \le 1} {E_{x\sim{p_{data}}(x)}}[F(\mathbf{x})] - {E_{z\sim{p_z}(z)}}[F(\underbrace {G(\mathbf{z})}_{\tilde x})] \\ +  &\lambda {E_{\hat x\sim{p_{\hat x}}(\hat x)}}\left[ {{{({{\left\| {{\nabla _{\hat x}}F(\hat x)} \right\|}_2} - 1)}^2}} \right] , \\ 
&\hat x = \rho \tilde x + (1 - \rho )x \quad and \quad 0 \le \rho  \le 1.
\end{split}
\end{equation}
\subsubsection{InfoGAN}
Chen et al. \cite{chenInfoGANInterpretableRepresentation} proposed a GAN framework for generating samples via the addition of conditional factor information to the generator noise input in a completely unsupervised manner. For example, if the goal is to generate the digits in the MNIST dataset, it would ideally be the case that the generative system has access to independent factor variables representing the digits' thickness and angle as part of the latent noise space. Such additional information can be denoted via ${c_1},{c_2},...,{c_L}$, and given that we are assuming that the latent variables are independent, the joint distribution can be written as $P({c_1},\,{c_2},...,{c_3}) = \prod\nolimits_{i = 1}^L {P({c_i})} $. In the following, $c$ is the concatenation of all ${c_i}$ variables. The InfoGAN cost function is, therefore:
\begin{equation}
\label{eq:10}
\mathop {\min }\limits_G \mathop {\max }\limits_D \,\,{L_{\rm{Info}}}(D,G) = L(D,G) - \lambda I(\mathbf{c;}G(\mathbf{z,c})),
\end{equation}
where $L(D,G)$  is as defined in Eq. (\ref{eq:1}), $I(\mathbf{c;}G(\mathbf{z,c})) = H(\mathbf{c}) - H(\mathbf{c}|G(\mathbf{z,c}))$, and $H$ is the entropy. Practically, the term $I(\mathbf{c;}G(\mathbf{z,c}))$  is hard to maximize directly as we would need to access to the posterior distribution $P(\mathbf{c}|\mathbf{x})$. However, a lower bound can be obtained variationally by defining a new structure $Q(\mathbf{c}|\mathbf{x})$, known as the auxiliary distribution. This lower bound is given as:
\begin{equation}
\label{eq:11}
\begin{split}
I(\mathbf{c;}G(\mathbf{z,c})) \ge& {E_{c \sim p(c),\,x \sim G(z,c)}}[\log Q(\mathbf{c}{\rm{|}}\mathbf{x})] + H(\mathbf{c})\\
=& {E_{x \sim G(z,c)}}[{E_{c' \sim p(c|x)}}[\log Q(\mathbf{c'}{\rm{|}}\mathbf{x})]] + H(\mathbf{c})\\ =& {L_I}(G,Q).
\end{split}
\end{equation}
Thus, the loss function can be represented as:
\begin{equation}
\label{eq:12}
\mathop {\min }\limits_{G,Q} \mathop {\max }\limits_D \,\,{L_{\rm{Info}}}(D,G,Q) = L(D,G) - \lambda {L_I}(G,Q).
\end{equation}
\subsubsection{Least Square GAN}
Mao et al. \cite{maoLeastSquaresGenerative} propose a novel loss function for addressing the vanishing gradient problem inherent in GANs (and deep learning generally) during the learning stage by replacing the cross-entropy loss by the least-squares loss. The new cost function for the discriminator can thus be written:
\begin{equation}
\label{eq:13}
\begin{split}
\mathop {\min }\limits_D \,L(D) =& \frac{1}{2}{E_{x\sim{p_{data}}(x)}}[{(D(\mathbf{x}) - 1)^2}]\\ +& \frac{1}{2}{E_{z\sim{p_z}(z)}}[{(D(G(\mathbf{z})))^2}].
\end{split}
\end{equation}
The loss function for the generator is similarly given as:
\begin{equation}
\label{eq:14}
\mathop {\min }\limits_G \,L(G) = \frac{1}{2}{E_{z\sim{p_z}(z)}}[{(D(G(\mathbf{z})) - 1)^2}].
\end{equation}
\subsubsection{Bidirectional GAN}
The Vanilla GAN maps from the latent noise space to the real data distribution; however, the standard GAN framework is not reversible and cannot map data to a latent layer representation. Donahue et al. \cite{donahueAdversarialFeatureLearning} consequently propose an unsupervised GAN framework, the Bidirectional GAN (BiGAN), which can map the real data distribution $\mathbf{x}$ to the latent noise domain $\mathbf{z}$ via a new Encoder structure that sits alongside the Generator and Discriminator. The objective function is defined as follows:
\begin{minipage}{\linewidth}
	\begin{equation}
	\label{eq:15}
	\begin{split}
	\mathop {\min }\limits_{G,E} \mathop {\max }\limits_D \,\,&L(D,E,G) = {E_{x\sim{p_{data}}(x)}}[{E_{z\sim{p_E}(.|x)}}[\log D(\mathbf{x,z})]]\\ &+ {E_{z\sim{p_z}(z)}}[{E_{x\sim{p_G}(.|z)}}[\log (1 - D(\mathbf{x},\mathbf{z})]].
	\end{split}
	\end{equation}
\end{minipage}
The above models are thus the principle GAN architecture variants applicable across the full range of machine-learning domains at the current time. We now look to the networking domain specifically.

\section{Applications Overview}
\label{sec:3}
We divide the use-cases of GANs in the literature into five main categories: mobile networks, network analysis, internet of things, physical layer, and cybersecurity. In each section, we shall briefly introduce the subject, focusing on the most common challenges faced by researchers prior to investigating how semi-supervised learning can address these challenges. In each case, we provide real-world examples. A summary of the reviewed applications is given at the end of the chapter.
\subsection{Mobile Networks}
With the rapid development of mobile networks over past decades, classical methods are unable to meet modern network demands. Therefore, researchers are looking into new methods, such as machine learning or novel optimization techniques, to keep up with the pace of demand and requirement changes. Zhang et al. \cite{zhangDeepLearningMobile} provide a detailed and broad-based review of deep learning methods and their application in wireless and mobile networks. Since our primary focus is on GANs, we review, in this subsection, their most notable applications in mobile networks.
\subsubsection{5G Network Slicing}
The fifth-generation mobile network is designed to meet the requirements of a highly mobile and fully connected society, enable automation in various industry sectors, and provide a viable infrastructure for `internet of things' applications. The traffic characteristics of these autonomously communicating devices are significantly different from human-made traffic. Hence, 5G networks are required to support very diverse functionality and performance requirements in order to offer the various services with reliability. In this context, the notion of {\em network slicing} is defined as a composition of network functions, network applications, and the underlying cloud infrastructure joined together so as to meet the requirements of a specific use case. Network slicing enables sharing a common infrastructure to deploy multiple logical, self-contained, and independent networks \cite{ordonez-lucenaNetworkSlicing5G}. With the aid of network slicing and creating various types of virtual networks, 5G can provide services with complex and dynamic time-variable resource requirements. Within this paradigm, however, different slices may have different resource demands, and the demands of a slice can be dynamic and vary during its operation time. Therefore, the need to predict user requirements concerning the different resources and the requirement of dynamically allocating these resources become crucial to the operation of 5G. 
\par
\begin{figure}
	\centering
	\includegraphics[width=\linewidth]{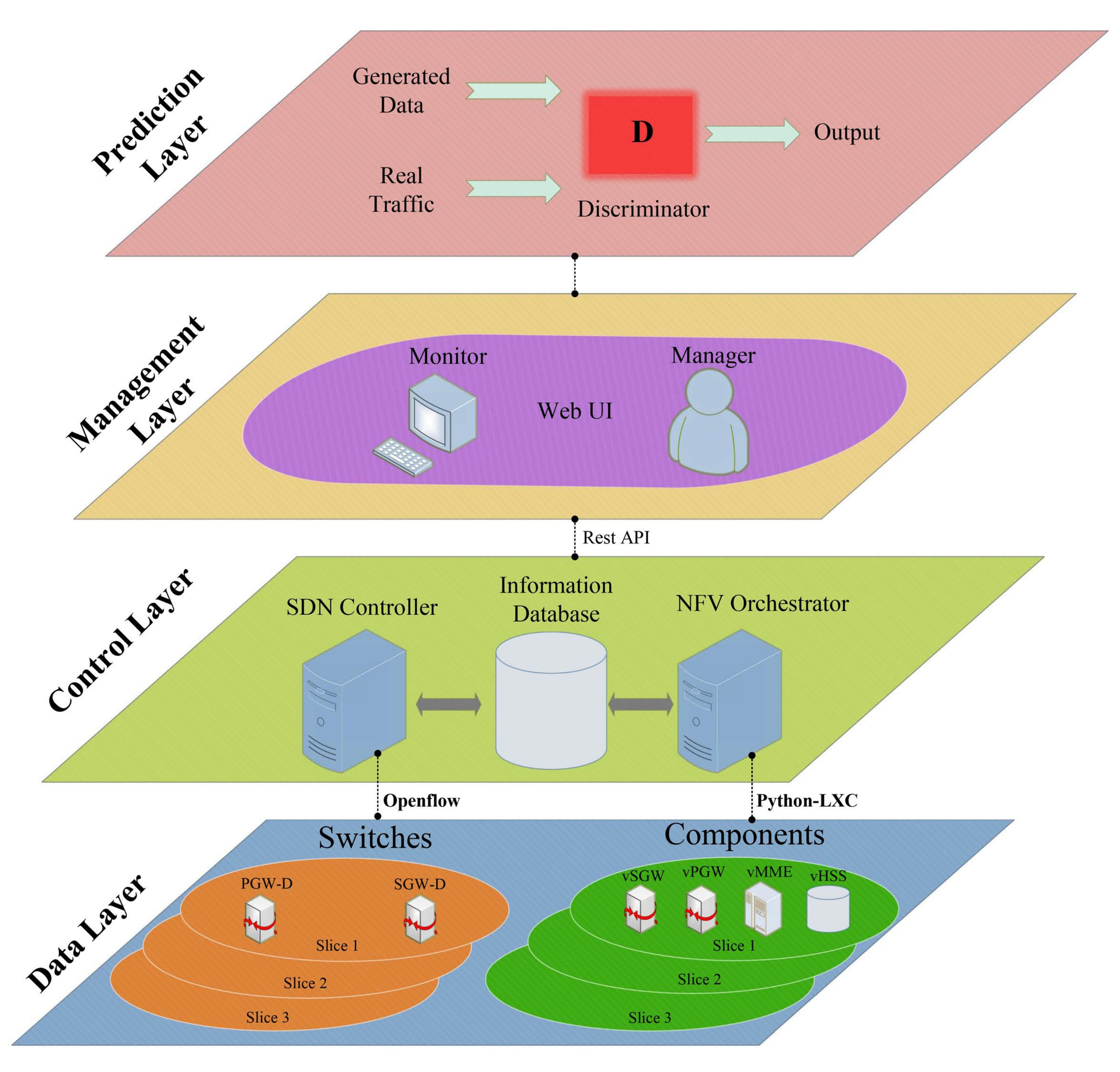}
	\caption{The architectural scheme of GANSlicing \cite{guGANSlicingGANBasedSoftware} for predicting slice demand by users. }
	\label{fig:3}
\end{figure}

 Gu and Zhang \cite{guGANSlicingGANBasedSoftware} proposed {\em GANSlicing}, a dynamic software-defined mobile network slicing framework for prediction of the resource demands of the internet of things (IoT) applications and also for improving the Quality of Experience (QoE) of users. GANSlicing aims to generate a global view of network resources that considers the physical and virtual capabilities of cellular networks to achieve more efficient utilization and allocation of resources to the network slices dynamically. A third use-case is the prediction of user demands for a variety of different resources via the underlying deep generative model. A GAN, which can generatively mimic an administrators' network operations, can also, in principle, predict traffic flow from historical information. Hence, GANslicing allows for slice demand to be forecast; so that the overall resource utilization is enhanced. The architecture of this model may be seen in Fig.\ref{fig:3}.
\par GANSlicing is implemented in two parts; service-oriented slicing and GAN-based prediction. Evaluation of the accuracy of network traffic prediction, and analysis of the performance of the slicing scheme (which the authors compared with tenant-oriented slicing, the most common scheme in mobile networks) indicates that GANSlicing can accept 16\% more requests with 12\% fewer resources in the same service request batch, hence improving the service acceptance ratio and enhancing overall service quality \cite{guGANSlicingGANBasedSoftware}. 
\subsubsection{Self-Organizing Networks}
Maintaining wireless and cellular networks' functionality and service provisioning while reducing low capital expenditure and operating expenditure has been challenging for both operators and service providers. Self-Organizing Networks (SONs) \cite{aliuSurveySelfOrganisation, imranChallenges5GHow} consist of a set of functions for automating the planning, configuration, management, and optimization of mobile networks. Since SON's primary function is to learn the parameters of a network and then optimize them, they inherently rely on data processing and intelligent decision making; consequently, learning algorithms and methods can be utilized to achieve this goal \cite{klaineSurveyMachineLearning, moysen4G5GSelforganized}. These methods include, but are not limited to, unsupervised learning \cite{palaciosUnsupervisedTechniqueAutomatic}, deep learning \cite{zhangSelfOrganizingCellularRadio}, Q-learning \cite{mwanjeCognitiveCellularNetworks} and Deep Q-learning \cite{mismarDeepQLearningSelfOrganizing}. However, most of these algorithms require gathering a large amount of real-case data to be effective. This becomes a challenge for many reasons; firstly, in many scenarios, the amount of data available for specific scenarios can often be limited, as such scenarios may be hard to reproduce. Secondly, gathering sufficient data can be costly or otherwise constrained, especially in scenarios where time is a limiting factor. Finally, a major challenge is data imbalance, given the infrequency of certain classes of events, which the SON must nonetheless take into account.
\par Recently, researchers have applied generative algorithms to overcome the aforementioned challenges, proposing various methods to improve SON performance and efficiency through synthetic data generation in order to increase the available quality of training data. For example, Zhang et al. \cite{zhangGenerativeAdversarialLearningBased} investigated the application of traditional classification algorithms for cell outage detection in SONs. Since cell outage is a relatively rare and low probability event, they typically constitute only a tiny fraction of total network measurement data. Due to this data imbalance, traditional learning algorithms will tend to construct a biased classifier. Consequently, the classifier output exhibits a skewness towards the majority class. The authors consequently proposed a novel cell outage detection scheme by combining Vanilla GAN with adaptive boosting (Adaboost). By utilizing GANs to generate synthetic data for the minority classes, they were able to correct this imbalance so that Adaboost can then be used to classify the re-balanced data and effectively detect cell outage.
\par Hughes et al. \cite{hughesGenerativeAdversarialLearning} presented a further application of GANs to SONs, seeking to augment Call Data Records obtained from a mobile operator. These data records exhibit two main features; call duration and start hour, for which the authors were able to generate synthetic tabular data. It is notable in this study that the reported difference in the variance of the generated and real data was more significant than that of the difference in mean values, a consequence of the fact that GANs often fail to generate realistic outlier values. They nonetheless achieved an accuracy improvement of 3.43\%.
\subsection{Network Analysis}
Network analysis is the act of collecting network data and analyzing it in order to improve the overall performance, reliability, and security of the network. These data usually consist of packets, log files, and configuration data. Computer and communication networks typically require real-time data delivery. However, much of this data is unstructured and unprocessed; consequently, the requirement for efficient tools and means of data analysis becomes critical. 
\par Since machine learning methods have the benefit of leveraging statistical patterns in data to perform analysis tasks with little pre-programming, they have received much attention in network analysis; one such common task is the analysis and control of communication networks. Machine learning methods, however, require large volumes of data to have acceptable performance in this context. Since networks tend to be well distributed, this large volume of data must be collected from several points in the network so that the learning algorithm can achieve a sufficiently global perspective on the network. Aho et al. \cite{ahoGeneratingRealisticData}, provided a novel approach to generate synthetic live traffic in order to improve the robustness of the learning algorithms applied to network analysis by applying several GAN variants. In particular, they utilized adversarial networks to generate network traffic similar to original samples. By comparing evaluation metrics for five different GAN architectures, Vanilla GAN, CGAN, LSGAN, WGAN, and WGAN-GP, they concluded that for network data generation, GAN and LSGAN are impractical, while WGAN and its variants offer significant performance gains.
\par Network traffic classification is typically one of the first steps in network analysis. In order to provide better service quality and also for management and security purposes, service providers must establish the different types of network application. Due to the massive volume of network data being transmitted, this task is usually automated. In general, there are two types of method applied to achieve this task; classification using payloads of packets and classification based on statistical analysis \cite{joseWANApplicationOptimization}. Researchers have consequently utilized various learning algorithms for traffic classification over the last few years \cite{shafiqNetworkTrafficClassification}. For security and user privacy purposes, many applications use network protocols such as SSL or VPN to encrypt their traffic. However, this encryption makes the analysis and classification of network data a challenging task. One of the major challenges in encrypted traffic classification is class imbalance, given that the majority of network traffic is regular and unencrypted traffic. In this context, Wang et al. \cite{wangFLOWGANUnbalancedNetwork} proposed FlowGAN, a method that uses GANs to generate synthetic traffic data for classes that suffer from low sample counts. They then used an MLP classifier to evaluate the effectiveness of their method, finding that tackling the class imbalance problem in this manner can indeed increase the performance traffic classifiers.
\par Li et al. \cite{liDynamicTrafficFeature} proposed FlowGAN (not to be confused with the FlowGAN proposed in \cite{wangFLOWGANUnbalancedNetwork}), a novel dynamic traffic camouflaging method to mitigate traffic analysis attacks and circumvent censorship. The idea behind this method is to use a GAN to learn features of permitted network flow (the target) and morph on-going censored traffic flows (the source) based on these features, in such a way that the morphed traffic is indistinguishable from the real flow. The authors utilized WGAN for the generator and WGAN-GP for the discriminator, evaluating the resulting method on more than 10,000 network flows, the data consisting of 6 features: outgoing packets, incoming packets, byte counts of outgoing packets, byte counts of incoming packets, cumulative bytes and the average interval between packets. Since traffic analysis attacks are principally a classification problem between different traffic data, the authors evaluated their method using area under curve and Indistinguishability under Classification Attack (IND-CA), defined in \cite{liDynamicTrafficFeature} as:
\begin{equation}
\label{eq:16}
IND - CA = \frac{{\left| {\Pr \left[ {\textrm{Priv}K = 1} \right] - 0.5} \right|}}{{0.5}},
\end{equation}
where $\textrm{Priv}K = 1$ if the attacker can distinguish between traffic flows.
\par Dynamics play a very significant role in the performance of most network systems; therefore, it is crucial to predict dynamics while performing network analysis. For instance, in an ad-hoc network, the dynamics of communication links make designing routing protocols challenging. Lei et al. \cite{leiGCNGANNonlinearTemporal} formulated the dynamics prediction problem of various network systems as temporal link prediction tasks, where abstracted dynamic graphs describe the system's behavior. They proposed a novel non-linear model, GCN-GAN, to predict these links in a weighted dynamic network. This model combines a Graph Convolutional Network (GCN) and a Long Short-Term Memory (LSTM) network with a GAN to improve representation learning and generate high-quality and plausible graph snapshots. By using GCN and LSTM as hidden layers of the generator network, the generator is able to predict the subsequent snapshots based on the historical topology of the dynamic graph.
\par Social network analysis is a particular subgroup of network analysis, {\em social tie prediction} being a quintessential problem in which social network operators attempt to suggest new connections or products to users based on their current social activity. Chen et al. \cite{chenTranGANGenerativeAdversarial} proposed TranGAN, a GAN-based transfer learning method for social tie prediction that seeks to uncover latent information in social networks. TranGAN, inspired by Triple-GAN \cite{liTripleGenerativeAdversarial}, in addition to the usual generator and discriminator structure, utilizes an additional Neural Network (NN) classifier for assigning labels to output samples from the generator. Although in this transfer learning scenario, the source and the target network are from different domains, the composite system is able to use information from the source network to improve the performance of the target network. The source network contains well-labeled social relationships; the labels of these relationships are missing in the target network, and hence the two are heterogeneous. 
\par As indicated, since most current network security systems utilize ML-based algorithms to perform network analysis, the data requirement needed to train these systems can become problematic. While simulations may be able to generate sufficient data, these are typically laborious and time-consuming to create. Xie et al. \cite{xieEffectiveMethodGenerate}, proposed utilizing existing network attack data generation tools augmented with data generated by WGAN. While simple and effective, this method is only able to generate continuous network features, since WGAN does not perform well in generating discrete features such as ``protocol type'' ``flag'' and ``service.''
\subsection{Internet of Things}
\subsubsection{Wireless Sensor Networks}
As an intermediate layer between wireless sensor networks (WSNs) and the end-user, middleware can provide a solution to various design issues, including security, heterogeneity, and performance scalability. One of the key challenges in WSNs is security, such that the confidentiality, authenticity, and integrity of data transmission from sensors can be guaranteed \cite{hadimMiddlewareWirelessSensor}. However, most middleware approaches cannot fully guarantee these security properties and protect the network from malicious attacks. Alshinina and Elleithy \cite{alshininaHighlyAccurateMachine} proposed a unique WSN middleware, powered by a GAN, for overcoming these design challenges while providing appropriate security measurement for handling large scale WSNs. This proposed method, in common with other GANs, consists of a generator and a discriminator. The generator creates fake data with similar attributes to the real data to confuse would-be attackers (the WSN does not need to generate fake data in this case so that power consumption can be significantly reduced). Correspondingly, the discriminator is tuned to distinguish real data and detect anomalies for further processing. 
\subsubsection{Indoor Localization and path planning}
With the exponential growth of smartphones and wearable technologies over the last decade, demand for location-based services has significantly increased. Currently, widely used localization services such as the global positioning system, while demonstrating excellent performance in ideal outdoor environments, can perform poorly in indoor or harsh outdoor environments. This may be caused by many phenomena: fading, shadowing, multipath, and a lack of line-of-sight \cite{wangCSIBasedFingerprintingIndoor}. For this reason, other techniques with high accuracies, such as fingerprinting, are generally used for indoor localization. Currently, due to its high availability and ease of access, WiFi-based fingerprinting is the most commonly used method for indoor localization \cite{homayounvalaNovelSmartphoneApplication}. The fingerprinting method consists in two phases; a training (offline) phase, where signal feature measurements are initially collected, and a test (online) phase, in which real-time signal properties are measured and compared against the offline phase data in order to provide an accurate estimation of the desired location. The two primary signal properties used for fingerprinting are Received Signal Strength (RSS) and Channel State Information (CSI) \cite{wangCSIBasedFingerprintingIndoor}. Machine learning and NNs have been widely used to learn probability features from these signal properties to perform localization.
\par However, such ML-based techniques face a crucial challenge in a lack of, or shortcoming in, manually-labeled data for training. Moreover, in many cases, data collection is costly and time-consuming (for instance, in crowd-sourcing, a large number of human participants are typically required to collect and annotate data via their mobile phones). Nabati et al. \cite{nabatiUsingSyntheticData} proposed to address this issue through the use of GANs to learn the underlying distribution of collected Received Signal Strength Indicator (RSSI) datasets in order to generate synthetic data and increase the amount available during the offline phase. Their proposed method, as well as reducing the cost of data collection, achieved identical benchmark accuracy with a lower real data requirement in a shorter time (they can use as little as 10\% of the real data with the associated reduction in data collection costs while achieving the same accuracy levels).
\par Li et al. \cite{liWaveletTransformDCGAN} converted collected CSI data using complex wavelet transforms in order to create feature maps that can extend the acquired fingerprint database through the use of the Wavelet Transform Feature DCGAN. This approach accelerates the convergence process during the training phase and is able to increase the variety of generated feature maps substantially. Experimental results indicated that the method can generate CSI data with improved diversity. The corresponding increase in the number of samples in the training set allows for significantly better localization accuracy, thereby demonstrating the proposed model's superiority to existing fingerprint database construction methods.
\par Mohammadi et al. \cite{mohammadiPathPlanningSupport} investigated path planning, proposing a GAN architecture to suggest accurate and reliable paths for differing applications, such as wayfinding for disabled or visually-impaired people. This approach utilizes a GAN to generate paths to a destination via collected user trajectory data. This involves two components: localization and wayfinding. The GAN structure uses a feed-forward NN at the input layer with four hidden layers and a hyperbolic tangent activation function as the generator; the discriminator is similarly a feed-forward NN. The localization aspect's training process constitutes a multi-label classification task, accomplished via the classification of RSSI values into location coordinates. A separate model was created for each class feature and trained on the common training dataset. The path classifier was evaluated separately as it required truth values to determine how well the model can identify classes. Experimental results suggested that paths generated by the model are more than 99.9\% reliable, with the path classifier able to classify the given path with around 99\% accuracy. 
\subsubsection{Human Activity Recognition}
\par Human Activity Recognition (HAR) has a significant role in many IoT applications such as smart houses, health care, and elderly monitoring. Within this domain, many approaches involve motion sensors, cameras, and WiFi signals. However, motion sensors, though accurate, may be expensive and impractical, and wearing them on the person proves detrimental to user well-being. Using image signals gathered from cameras is also frequently unviable, given that they are unable to work in darkness or non-line-of-sight settings. This often leaves WiFi signals as a very useful supplementary, or even the best, HAR option. However, straight WiFi-based HAR, due to the low-resolution and limited sensing capability of RSSI measurements, is often unable to achieve fine-grained HAR. This has prompted recent studies to propose CSI measurements as a way to achieve recognition performance \cite{liuWirelessSensingHuman}.
\par Aiming to leverage WiFi signal's pervasiveness without the requirement for specialized equipment, Yousefi et al. \cite{yousefiSurveyBehaviorRecognition} applied CSI to the problem of recognizing seven distinct human activities. Each action has a distinct pattern, and the distinct motions should therefore have differing effects on the CSI. However, owing to the low signal-to-noise ratio, raw CSI measurements are not in themselves sufficiently representative of these different human activities. Rather than hand-craft discriminative features, the authors propose an LSTM-based approach to learn representative features to encode the temporal information. However, Moshiri et al. \cite{moshiriUsingGANEnhance} indicated that it is difficult to collect adequate labeled data for training the proposed LSTM model, thus they proposed a semi-supervised GAN-based solution instead. Hence, they generated an augmented dataset with the same statistical features as the real data by presenting 50\% of each activity class to the GAN model, with their proposed model improving classification accuracy and scaling-down the Log Loss to give an overall accuracy improvement of 3\%. 
\par Xiao et al. \cite{xiaoCsiGANRobustChannel} applied leave-one-subject-out validation to CSI-based activity recognition to address the performance degradation problem. Using a GAN-based framework, termed CsiGAN, they conducted experiments on two CSI-based behavior recognition datasets; SignFi, which includes CSI traces concerning sign language gestures, and FallDefi, which includes CSI traces concerning a range of typical human activities including falling, walking, jumping, picking-up, sitting-down, and standing-up. The semi-supervised GAN used in this paper extends the standard GAN discriminator by adjusting the number of probabilistic outputs from $k+1$ into $2k+1$ (where $k$ is the number of categories), which helps in obtaining the correct decision boundary for each category. They also proposed a manifold regularization method to enhance classification performance by stabilizing the learning process.
\begin{figure}
	\centering
	\includegraphics[width=\linewidth]{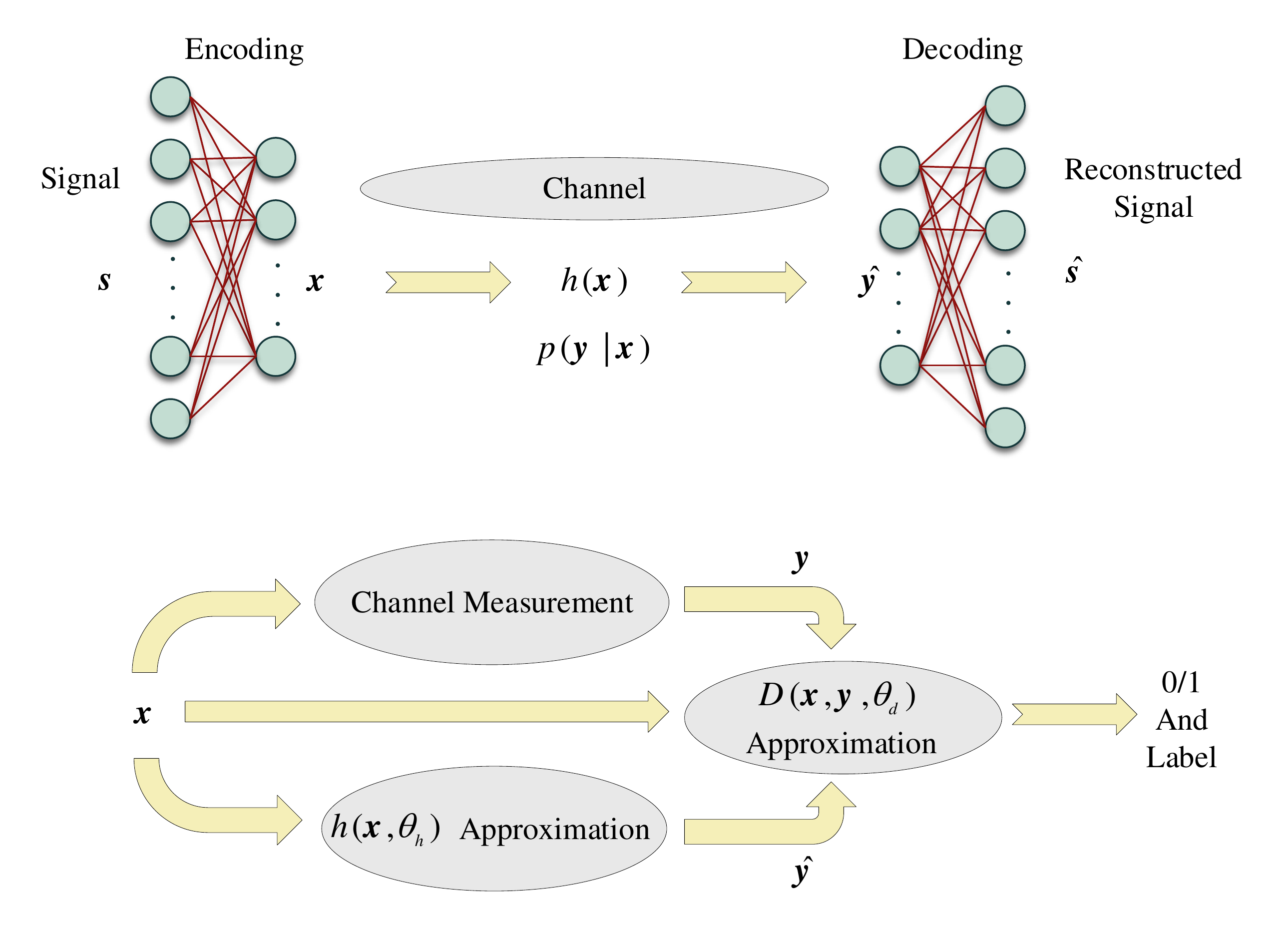}
	\caption{(a) Modeling an end-to-end communication system with autoencoders, as proposed in \cite{osheaIntroductionDeepLearning}.
		(b) approximating the stochastic channel function using GANs.}
	\label{fig:4}
\end{figure}
\subsection{Physical Layer}
Deep learning has been broadly adopted in physical layer communications, especially for signal-related processes, including encoding, decoding, modulation, and equalization. It has shown exceptional capabilities in removing various constraints of existing communication systems, such as in wireless channel modeling. Traditional channel modeling methods are exceedingly complex, unique to the channel environment, and are typically unable to model the channel's key stochastic properties. Using NNs, however, allows us to overcome these drawbacks. For instance, an end-to-end communication system can be represented as an autoencoder as shown in Fig.\ref{fig:4}(a). This approach consists of an encoder encoding symbols into a distinct transmitted value, a stochastic channel model, and a decoder network which seeks to estimate the transmitted symbols from the received samples and outputs a probability distribution over all possible decoded messages \cite{osheaIntroductionDeepLearning}. 
\par O'Shea et al. \cite{osheaPhysicalLayerCommunications} proposed a method for physical layer modulation and coding for the communication system that uses adversarial learning. They thus employed GANs with channel autoencoders to approximate the channel's response and learned an optimal scheme under certain performance metrics. This approach, which they termed Communications GAN, utilizes two separate networks for encoding and decoding, both consisting of a fully connected layer with ReLu activation. Using the mean squared error as a measure of channel loss, normalization and noise interpolation were the main focus of training. Their results suggested that by learning a channel function approximation and an encoder/decoder layout, robust performance without explicit prior implementation can be achieved. Such a system can learn directly on unseen physical channels, an approximation of these channels being sufficient for adapting the encoder and decoder networks.
\par The approach described above, although model-free, assumes that the channel model function is differentiable. Otherwise, gradients could not be computed during the back-propagative training of the network. However, since we do not have access to an exact channel model in reality, this has to be estimated. Analytic channel models can only express a limited number of wireless channel effects (such as interference, propagation, distortion, noise, and fading); this is because expressing non-linear effects is laborious due to their complexity and high number of degrees of freedom. O'Shea et al. \cite{osheaApproximatingVoidLearning} extend their approach so as to represent a broad range of stochastic channel effects with a high degree of accuracy. The channel model is a stochastic function; hence it can be modeled as a conditional probability $p(y|x)$. Similarly, the channel approximation network can also be modeled as a conditional probability distribution, $p(\hat y|x)$ where $\hat y$ and $y$ represent synthetic and real samples, respectively, and $x$ is the channel input. The goal here is to minimize the distance between $p(y|x)$ and $p(\hat y|x)$ such that the model approximation becomes more accurate. This setup, as shown in Fig. \ref{fig:4}(b), is achieved via a new discriminator network (the authors asserted that this task could also be performed using a WGAN in order to improve training stability).
\par In this vein, Ye et al. \cite{yeChannelAgnosticEndtoEnd} proposed an end-to-end channel-agnostic communication model that can be applied to more realistic time-varying channels. They utilized CGAN to model the conditional distribution $p(y|x)$, where the transmitter's encoded signal constitutes the conditioning information. By adding pilot information to the conditioning information, the system is able to generate more specific samples and estimate the CSI more accurately. This approach enables end-to-end learning of a communication system without prior information regarding the channel. In other words, by training a conditional GAN, the transmitter, receiver, and end-to-end loss can be well-optimized in a supervised manner. The authors initially applied their method to the Additive White Gaussian Channel (AWGN) channel, in which the output, $y$, is a summation of the input signal, $x$, with Gaussian noise, $w$, such that $y = x + w$. In this case, the conditioning information is the encoded signal from the transmitter since channel estimation is not required. In a further experiment, Rayleigh fading channels were studied, for which the output is given via ${y_n} = {h_n}{x_n} + {h_n}$ where ${h_n}\sim CN(0,1)$; since the channel is time-varying, additional conditional information needs to be appended to the channel receiver. The authors assert that the system may be further extended to other types of channels, beyond those of AWGN and Rayleigh fading.
\par
\begin{figure}
	\centering
	\includegraphics[width=\linewidth]{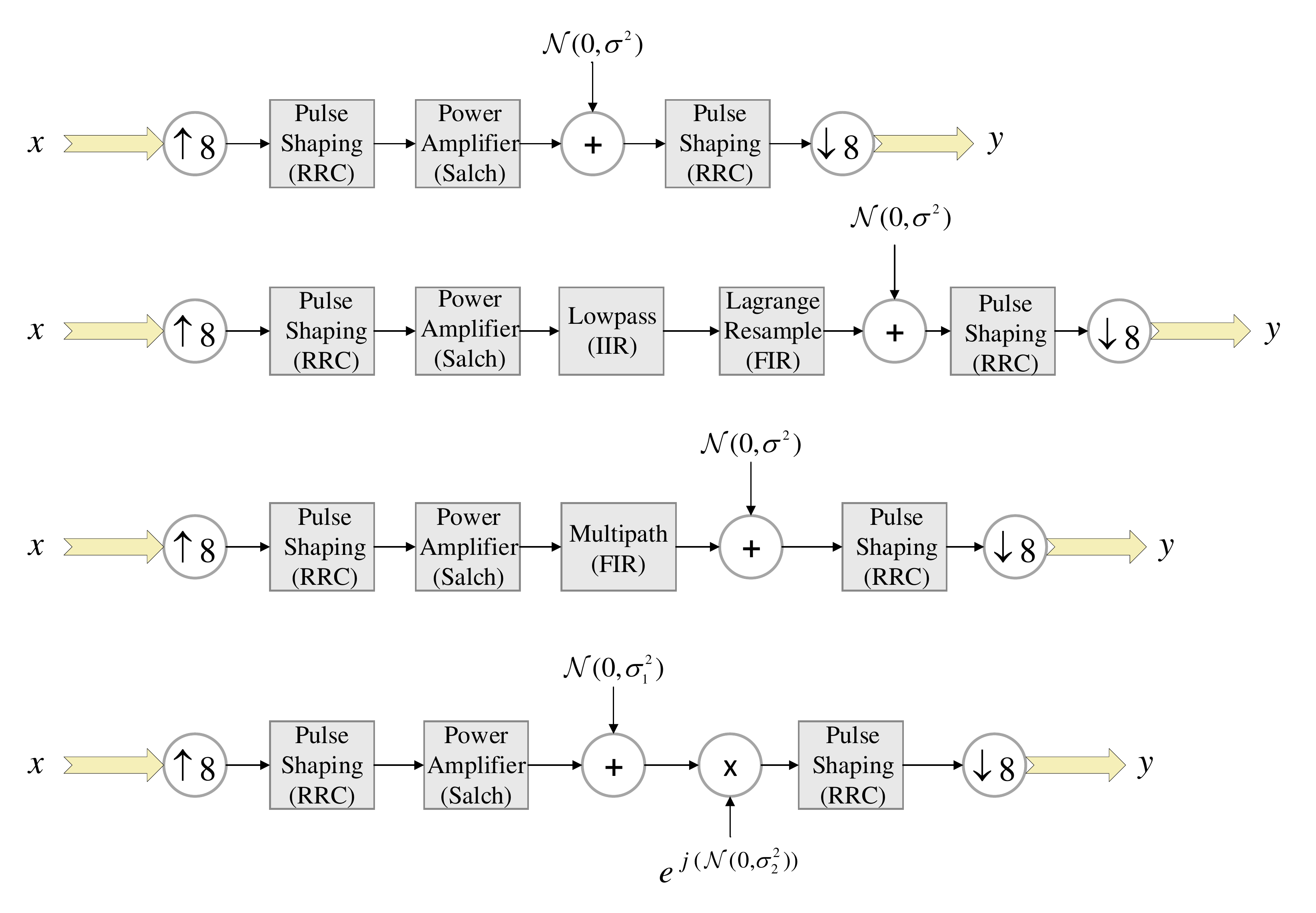}
	\caption{The different models tested in  \cite{smithCommunicationChannelDensity}.}
	\label{fig:5}
\end{figure}

 Autoencoder-based communications systems have become pervasive in the research community. In most cases, researchers use the encoder as a transmitter that maps messages to symbols. However, the problem is that the gradients of the physical channel are often obscure, and this circumstance prevents the transmitter network from receiving updates during training. One solution is to approximate the channel response using a NN, such that the NN can act as a substitute for the physical channel during the training process. An early example of this method was presented in \cite{osheaApproximatingVoidLearning}, where the authors used a GAN to approximate a stochastic channel that includes non-linear distortions and non-Gaussian statistics. Results indicated the utility of the GAN; however, the Probability Density Function (PDF) of the learned channel model differs from the simulated channel's PDF, and the channel is further assumed to have no memory effect. Smith and Downey \cite{smithCommunicationChannelDensity}, inspired by BicycleGAN \cite{zhuMultimodalImagetoImageTranslation}, evaluated a novel GAN architecture for learning nonlinearities, memory effects and non-Gaussian statistics. Their research focuses on channels that contain a combination of non-linear amplifier distortion, pulse shape filtering, inter-symbol interference, frequency-dependent group delay, multipath, and non-Gaussian statistics. They compared the marginalized PDFs of the channel with a trained generator. Carrying out experiments on four different channels, as shown in Fig.\ref{fig:5}, their results suggest that the proposed model is capable of generating high-accuracy approximations of the channel.
\par Yang et al. \cite{yangGenerativeAdversarialNetworkBasedWirelessChannel} proposed an alternative channel modeling framework utilizing GANs. The GAN in question was trained using raw channel measurement data, for which the Nash equilibrium is the convergence of a minimax game between the generator and discriminator. The generator thus learns the distribution of channel impulse responses and generates synthetic samples. Once the equilibrium point is reached, the generator can be extracted as the target channel model. To evaluate the suggested framework's performance and verify the effectiveness of the method, they estimated the channel response of an AWGN channel, comparing the PDF of the learned channel model with the real AWGN channel. The authors suggested that this framework could, in principle, be extended to large-scale channels, such as multiple-input/multiple-output.
\par Zhao et al. \cite{zhaoClassificationSmallUAVs} studied the detection, tracking, and classification of Unnamed Aerial Vehicles (UAVs). They used an oscilloscope and antenna to collect wireless signals in an indoor environment with a sample rate of 20 GS/s. For the corresponding outdoor environment, they utilized a universal radio software peripheral that uses IEEE 802.11g for communication with 20 MHz bandwidth (since the bandwidth of the oscilloscope is around 2.5 GHz). They then analyzed the UAVs' wireless signal features via modified principal component analysis for dimension reduction. They hence used wireless signals to detect UAVs in a manner that does not depend on the size, line of sight, protocol standardization, or level of forensic tool support. Leveraging the high recognition rate of the ACGAN and the theory of WGAN, they then proposed a more stable model variant, AC-WGAN, that exhibits a classification rate greater than 95\% in indoor environments. Training samples were introduced to both the discriminator and generator with the negative loss updated by ascending a stochastic gradient. Testing samples were then imported into the discriminator, and the taxonomy of the signals was classified based on the value of the negative loss. This model is reportedly capable of detecting UAV wireless signals in outdoor environments from several hundred meters away.
\subsubsection{Cognitive Radio}
Cognitive radio is a concept that aims to overcome the limitations of wireless channels by making radios programmable, dynamically configurable, and capable of learning and adapting to minimize user interference and enhance overall performance. Using machine learning techniques enables cognitive radios to learn and make decisions without the need for explicit programming. Learning methods in cognitive radio have been widely surveyed in the literature, for example, \cite{clancyApplicationsMachineLearning, bkassinySurveyMachineLearningTechniques}. In general, cognitive radio problems can be divided into two categories: classification (e.g. for spectrum sensing or modulation recognition \cite{madhavanSpectrumSensingModulation}), and decision-making (such as in power control or adaptive modulation \cite{clancyApplicationsMachineLearning}). Thus far, researchers have applied GANs only in the former category, and the applicability of GANs for decision-making problems is yet to be determined. 
\par As indicated, machine learning is useful in automating cognitive radio functionalities by offering a means of reliably extracting and learning the intrinsic spectrum dynamics. There are two challenges in this task: firstly, the machine learning algorithm requires a significant amount of data in order to capture multiple channels and emitter characteristics to train the classifier. Secondly, the wireless channel is highly dynamic, and consequently, as the spectrum varies, training data previously identified for one spectrum environment cannot be reused in any such altered environment. To address these challenges, Davaslioglu and Sagduyu \cite{davasliogluGenerativeAdversarialLearning} proposed Spectrum Augmentation/Adaptation via a novel GAN, SAGA, which utilizes CGAN to generate synthetic training data. SAGA thus aims to leverage training data augmentation and domain adaption to improve classifier accuracy and enable the classifier to operate in novel, unseen environments. The authors asserted that SAGA can further be extended to wideband spectrum sensing, in which multiple channels are present, by applying training data augmentation and adjustment.
\par Existing methods for modulation recognition present in the literature are mainly based on deep learning, given that modulation recognition is inherently a classification problem. In this context, Li et al. \cite{liRadioClassifyGenerative} proposed Radio Classify GAN (RCGAN), a novel end-to-end semi-supervised framework for modulation recognition. By utilizing DCGAN with a cost function and replacing the last layer of the discriminator network with a softmax function, they were able to classify radio signals presented in the form of complex time-domain vectors to achieve modulation recognition. Furthermore, their experimental results suggested that this novel approach can improve overall recognition accuracy even when the signal-to-noise ratio is under 4dB.
\par Before the connection between two transceivers being established, wireless signals are required to be authenticated at the physical layer. One form of wireless attack that targets this task is {\em signal spoofing}. In this type of attack, the adversary aims to impersonate a legitimate transmitter. This is usually done to bypass authentication systems or primary user emulation in cognitive radio, in which a secondary user mimics a primary user to occupy more of the spectrum. Shi et al. \cite{shiGenerativeAdversarialNetwork} used GANs to spoof wireless signals by generating and transmitting false signals. They assumed that a deep classifier is used at the receiver to predict the intentional transmission such that, if there was no attack, a pre-trained deep learning-based classifier could discern signals. In a standard spoofing system, such as a replay attack, the probability of success with respect to the deep classifier stays confined. However, the authors show that a GAN-based spoofing attack, in which generated signals are transmitted for the spoofing attacks, has the potential to enhance the success probability of wireless signal spoofing even when a NN classifier has been used as a defense mechanism. 
\par Erpek et al. \cite{erpekDeepLearningLaunching} investigated the security aspects of cognitive radio in case of wireless jamming attacks. They described different types of wireless jamming attacks and applied adversarial learning to design both the jamming attack and also an appropriate defensive scheme. Jamming is severely dependent on the training data to give appropriate information regarding the channel status. However, when a jammer can only collect limited data, its performance drops significantly. Therefore, they proposed that by using CGAN effectively, they can overcome this performance drop and shorten the learning period, hence making the exploratory jamming attack more efficient. Their results showed that by using just ten samples instead of the full 500 and utilizing this to generate synthetic data with CGAN, their misdetection probability and false alarm rate stays within 0.19\% and 3.14\% of that of the original data, which significantly reduces the overall time and cost of gathering data.
\par {\em Covert communication} is a novel communication paradigm with a low chance of being detected or intercepted \cite{soltaniCovertWirelessCommunication}. Since the two players in a GAN contend against each other to achieve the Nash equilibrium, GANs can be used to model an adversarial game between a legitimate user and a watchful warden in covert communications. In such a scenario, the generator, acting as a legitimate user, can be utilized to generate a covert transmit power allocation solution. Simultaneously, the discriminator can act as the warden and attempt to figure out the covert messages \cite{liaoGenerativeAdversarialNetwork}. Liao et al. \cite{liaoGenerativeAdversarialNetwork} considered a cooperative cognitive radio network, which benefits from a secondary transmitter acting as a relay. In this case, this secondary transmitter covertly transmits private information while being supervised by the primary transmitter. This scheme, termed GAN-based Power Allocation (GAN-PA), aims to seek a balance between the covert rate and the detection error. Experimental results suggested that the proposed scheme achieves near-optimal performance in the presence of minimal network status.
\subsection{Cybersecurity}
Cybersecurity is a complex of technologies, practices, and processes aimed to protect computers, networks, devices, and data from arbitrary cyber-attacks, unauthorized access, or malicious activity. Deep Learning has recently been widely applied to cybersecurity systems, for instance, in \cite{buczakSurveyDataMining, wickramasingheGeneralizationDeepLearning, al-garadiSurveyMachineDeep, bermanSurveyDeepLearning, tangDeepLearningApproach}. Because of their nature and capacity to alleviate the challenge of imbalanced datasets, GANs have been identified as having high potential in security and adversarial applications. We hence review GAN applications in Intrusion Detection Systems (IDSs) \cite{usamaGenerativeAdversarialNetworks, salemAnomalyGenerationUsing, linIDSGANGenerativeAdversarial, seoGIDSGANBased}, malware detection \cite{aminAndroidMalwareDetection, kimMalwareDetectionUsing}, detection of rogue Radio Frequency (RF) transmitters \cite{royDetectionRogueRF}, malware adaption/improvement \cite{rigakiBringingGANKnifeFight, huGeneratingAdversarialMalware, kawaiImprovedMalGANAvoiding}, black-box Application Programming Interfaces (API) attacks \cite{papernotPracticalBlackBoxAttacks} and other cybersecurity applications such as password guessing \cite{hitajPassGANDeepLearning} and credit-card fraud detection \cite{chenCreditCardFraud,wangFraudulentDataSimulation,sethiaDataAugmentationUsing}.
\subsubsection{IDSs, Malware Detection, and Security Systems}
IDSs play an essential role in maintaining network security. Their main task is to monitor network traffic and provide a defense against unusual and malicious traffic. Usama et al. \cite{usamaGenerativeAdversarialNetworks} pointed out the vulnerability of IDSs towards adversarial examples and generative adversarial attacks in order to evade IDS detection. The GAN-based attack they envisaged adds perturbation to traffic features so that the IDS would be unable to detect it as malicious traffic. They further expanded this idea, proposing a GAN-based defense to increase the robustness of IDSs to this kind of attack.
\par Similarly, a novel framework based on WGAN called IDSGAN is proposed in \cite{linIDSGANGenerativeAdversarial} to generate adversarial attacks capable of evading IDS detection. This framework consists of a generator, a discriminator, and a black-box IDS based on the fact that the IDS structure is unknown to the attackers in a real-case scenario. IDSs are also an essential part of in-vehicle networks. However, such IDSs require very high accuracy as any error in detection may seriously endanger driver and passengers' safety. Seo et al. \cite{seoGIDSGANBased} proposed a GAN-based IDS (GIDS) for detecting unknown attacks using normal data; since GIDS constitutes a pre-trained model with two discriminators, one to detect known attacks and the other to detect unknown attacks, it may be applied as a real-time intrusion detector with excellent performance reported.
\par Data imbalances can occur in anomaly-detection problems since, in typical datasets, the anomalous instances are rare compared to the normal class. A learning model trained on such a dataset will naturally favor the majority class and perform poorly. GANs show potential in addressing such imbalances and are thus valuable in scenarios where gathering or generating anomalies is costly and time-consuming. Salem et al. \cite{salemAnomalyGenerationUsing} investigated this in the domain of Host-based IDSs (HIDSs), where normal data is abundant as compared to anomalies. They made use of the ADFA-LFD dataset \cite{creechSemanticApproachHostBased}, and after converting the numeric data to images, utilize a Cycle-GAN model to learn the transformation between normal and anomaly data in order to generate anomalies. In this way, by creating a framework to transform normal data into anomalies and adding these anomalies to the original dataset, they can significantly increase the performance of HIDSs.
\par Malware is defined as any application that exhibits malicious behavior, such as viruses, worms, Trojans, and ransomware. In contrast, benign applications are legitimate programs that are not harmful and perform their intended actions with the full acknowledgment of the user. Generally, the primary purpose of any anti-malware application is to distinguish between malware and benign applications. Amin et al. \cite{aminAndroidMalwareDetection} investigated Android operating system malware, proposing a novel malware detection method that uses GANs with LSTM hidden layers (LSTM-GAN) in both the generator and discriminator. The system's input consists in binary opcode sequences extracted from different applications, represented as 1D tensors. After training a sufficient number of epochs, the discriminator may then be used as a malware detector. Kim et al. \cite{kimMalwareDetectionUsing} suggested pre-training the generator using an autoencoder to overcome instability during training. This model, called transferred GAN, was then used to detect malware, particularly zero-day attacks for which only a tiny amount of data is available.
\par Roy et al. \cite{royDetectionRogueRF} addressed the problem of identifying rogue RF transmitters with the help of adversarial networks. All transceivers display a unique unwanted in-phase and quadrative imbalance (IQ imbalance) \cite{chia-lingliuImpactsImbalanceQPSKOFDMQAM}. Therefore, by exploiting the IQ imbalance in RF transmitters, they proposed that Vanilla GAN can learn and generate unique features, using them as fingerprints to identify and classify transmitters. Thus, the generator model generates fake signals to spoof the transmission of known transmitters, and the discriminator detects these rogue transmitters. After training this model using over-the-air data collected from trusted transmitters, the discriminator can detect fake transmitters with about 99.99\% accuracy.
\par Shin et al. \cite{shinAndroidGANDefendingAndroid} proposed a GAN-based model to defend against attacks aimed at the Android pattern-lock system. They suggested a Vanilla GAN based anomaly detection paradigm utilizing only single-user data that would generate a large amount of synthetic data, which may then be treated as the potential attacker during the training phase. To enhance stability during training, they used a Replay Buffer to generate high-quality synthetic data. Expanding this idea further, they introduced a multi-modal network that uses two-touch features (trajectory and pressure), resulting in a robust anomaly detector that proves effective against this form of attack.
\subsubsection{Malware Adaption and Improvement}
In contrast to the foregoing, GANs may also be used to increase the robustness of malware. Rigaki and Garcia \cite{rigakiBringingGANKnifeFight} utilized GANs to enable malware to adapt to changing conditions and hence become harder to detect. Their proposed method used LSTM-GAN to learn the features of benign application traffic flow (in this case, Facebook chat) in order to imitate them. After convergence, the malware then connects to the generator and by using the output parameters, adapts its traffic accordingly. Experimental results found that with just 217 real-case network flows trained for 400 epochs, the blocked malware percentage dropped to zero.
\par Malware usually possesses either very little or no information about the structure and parameters of IDS models, and hence the target system may be treated as a black-box. However, it is possible to predict what features these systems use. Hu and Tan \cite{huGeneratingAdversarialMalware} proposed a novel algorithm, MalGAN (depicted in Fig. 6), which generates complex and flexible adversarial malware examples to attack and attempt to fool the black-box detector. Once the generator is trained on sufficient data, it can generate adversarial examples having probability distributions far from that which the black-box detector is trained on, therefore classified as benign. Experimental results showed that, with a NN substitute detector as the black-box, almost all of the adversarial examples generated by MalGAN bypass the detection algorithm.
\par However, MalGAN does suffer from few issues; first, malware detectors must be built internally. Secondly, the feature dimension is reduced to just 128, meaning that not all original malware features are covered. Thirdly, both MalGAN and the detector use the same features. Lastly, multiple malware instances are required to train MalGAN \cite{kawaiImprovedMalGANAvoiding}. Kawai et al. \cite{kawaiImprovedMalGANAvoiding} hence suggested a number of enhancements, proposing Improved MalGAN to further evade detection by adding benign features to the original malware. They also addressed the issue of the high quantity of generated data required to avoid detectors by adding a loss calculation layer to the generator.
\par Black-box API attacks \cite{papernotPracticalBlackBoxAttacks} constitute a set of exploratory attacks that can be launched to learn proprietary information such as underlying training data, learning algorithms, and hyperparameters, without any prior knowledge. Service providers usually limit the number of calls each user can make to the API to prevent such attacks. Consequently, the amount of training data the attacker obtains will necessarily be trivial. Shi et al. \cite{shiGenerativeAdversarialNetworks} suggested implementing these attacks with the aid of adversarial networks, demonstrating that even with a minimal quantity of training samples, GANs are successful in attacking.
\begin{figure}
	\centering
	\includegraphics[width=\linewidth]{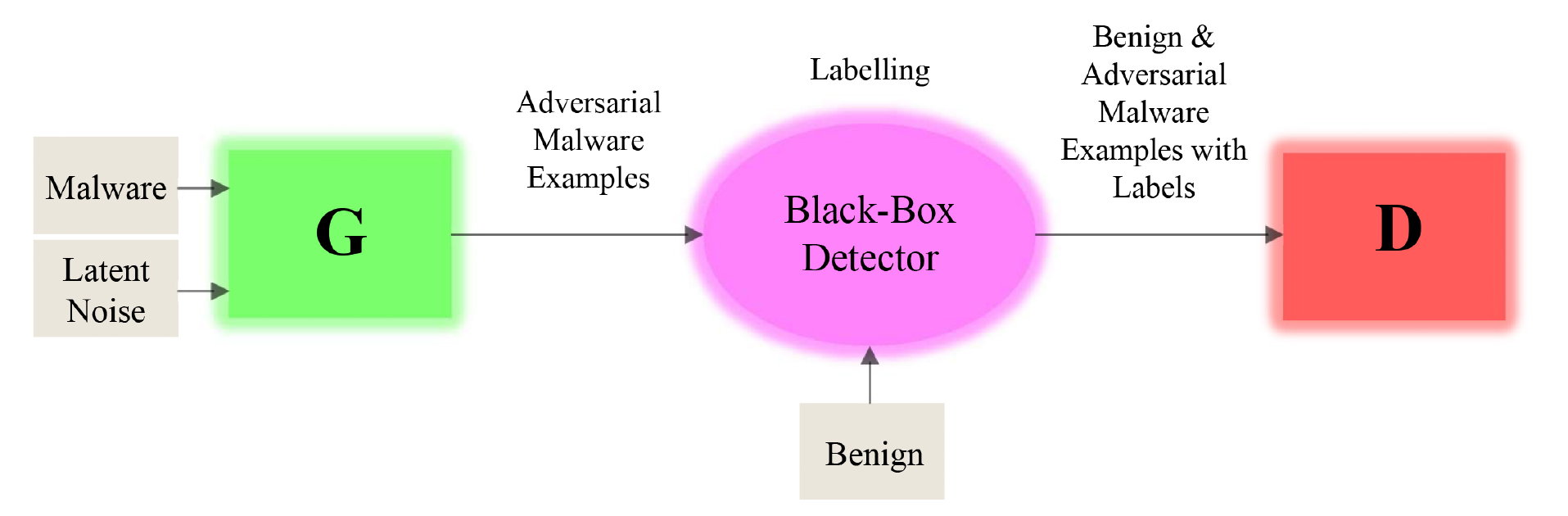}
	\caption{MalGAN architecture \cite{huGeneratingAdversarialMalware}.}
	\label{fig:6}
\end{figure}
\subsubsection{Other Security Applications}
Credit card fraud detection with machine learning requires a balanced dataset consisting of both standard and fraudulent transactions. However, in real-world scenarios, fraudulent transactions are far rarer than standard ones. This motivates the use of adversarial networks. For instance, \cite{chenCreditCardFraud, wangFraudulentDataSimulation} addressed this challenge from a variety of perspectives. Sethia et al. \cite{sethiaDataAugmentationUsing} applied Vanilla GAN, WGAN, RWGAN, MAGAN \cite{wangMAGANMarginAdaptation}, and LSGAN, with results suggesting that RGWAN provides the best performance, and vanilla GAN the worst due to mode collapse. Chen et al. \cite{chenCreditCardFraud} proposed a bespoke solution that uses a Sparse Autoencoder (SAE) to map regular transactions into a compact vector space with the GAN then seeking to generate synthetic standard transactions as well as identifying whether a given transaction is fraudulent or not. Hence, the generator generates fake standard transactions, and the discriminator attempts to distinguish between the generated data and the hidden representations learned from standard transactions by SAE. The final trained discriminator can then be used to detect fraudulent transactions. Wang et al. \cite{wangFraudulentDataSimulation} also provided a data enhancement model based on adversarial networks called SGAN in order to generate synthetic fraudulent data. 
\par Despite passwords being one of the most popular authentication methods, various password database leaks have demonstrated that users frequently choose simple passwords composed of common strings and numbers. Password guessing tools can hence be used as weak-password identifiers with passwords being stored in hashed form. State of the art password guessing tools, such as HashCat and John the Ripper, can check very large numbers of passwords per second against these password hashes. While such methods are very successful in practice, developing and testing new rules and the associated reconnaissance of user habits is time-consuming. To address this, Hitaj et al. \cite{hitajPassGANDeepLearning} proposed PassGAN, an approach that seeks to replace human-generated password rules with theory-grounded machine learning algorithms. PassGAN, based on WGAN-GP, exploits an adversarial network to learn the distribution of real passwords from actual password leaks to generate superior password guesses. Experimental results suggested that the number of matches increases steadily with the number of generated passwords. PassGAN is able to guess between 51\% and 73\% of new unique passwords than comparable tools, matching passwords that were not generated by any password tools. The authors indicated that training PassGAN on a larger dataset would need more complex NN structures, and consequently, a more comprehensive training regime; it would also be required to produce a larger quantity of passwords compared to other tools. However, they assert that these costs are negligible concerning the benefits of the method.
\par In summary, Table \ref{tab:3} provides a brief overview of the applications reviewed above with appropriate taxonomic categories appended. We highlight the main problems and challenges that the GANs in question aimed to overcome and the specific models used to achieve this task. By inspection of the table, it may be seen that GANs have, to date, principally been used within the various cybersecurity and physical layer fields. Most of the work done includes applications with limitations of data gathering or class imbalance within the dataset. However, there are also a few novel works that utilize the discriminator network to achieve classification tasks. 

\begin{table*}[!t]
	\centering
	\renewcommand{\arraystretch}{1.9}
	\caption{List of network-related research papers that utilize GANs, along with their respective categorization.}
	\label{tab:3}
	\resizebox{\textwidth}{!}{%
		\begin{tabular}{|c|c|c|c|c|}
			\hline
			\textbf{Class}                                 & \textbf{Reference}                                                                      & \textbf{Application}                        & \textbf{Problem}                                                                                                                                                      & \textbf{Used Model}                                                                                                \\ \hline
			\multirow{4}{*}{\textbf{Mobile   Networks}}    & Gu and   Zhang \cite{guGANSlicingGANBasedSoftware}                                      & Network   Slicing in 5G                     & \begin{tabular}[c]{@{}c@{}}Dynamicity of   demands in network slicing\\ Predicting resource requirements of different users\end{tabular}                       & Vanilla GAN                                                                                                         \\ \cline{2-5} 
			& Zhang et al. \cite{zhangSelfOrganizingCellularRadio}                                    & Cell outage   detection                     & Imbalanced   cell outage data in cellular networks                                                                                                                    & \begin{tabular}[c]{@{}c@{}}Vanilla GAN and Adaboost\end{tabular}                                            \\ \cline{2-5} 
			& Hughes et al. \cite{hughesGenerativeAdversarialLearning}                               & Generating   synthetic CDR                  & Requirement   of large amount of data                                                                                                                                 & Vanilla GAN                                                                                                        \\ \hline
			\multirow{7}{*}{\textbf{Network Analysis}}  & Aho et al. \cite{ahoGeneratingRealisticData}                                           & Generating   real network traffic           & Requirement   of a large volume of data for network analysis tools                                                                                                   & \begin{tabular}[c]{@{}c@{}}Vanilla GAN,\\LSGAN,\\EBGAN,\\WGAN,\\WGAN-GP\end{tabular}                       \\ \cline{2-5} 
			& Wang et al. \cite{wangFLOWGANUnbalancedNetwork}                                         & Traffic   classification                    & Class imbalance in the classification of encrypted traffic                                                                                                            & Vanilla GAN                                                                                                           \\ \cline{2-5} 
			& Li et al. \cite{liDynamicTrafficFeature}                                               & Traffic   camouflaging                      & Mitigating   traffic analysis attack and circumventing censorship                                                                                                     & WGAN and WGAN-GP                                                                                                        \\ \cline{2-5} 
			& Lei et al.   \cite{leiGCNGANNonlinearTemporal}                                         & Network link   prediction                   & Dynamics in   network systems                                                                                                                                         & Vanilla GAN                                                                                                            \\ \cline{2-5} 
			& Chen et al.   \cite{chenTranGANGenerativeAdversarial}                                  & Social tie   prediction                     & Lack of   annotations in social network links and connections                                                                                                         & Triple-GAN                                                                                                            \\ \cline{2-5} 
			& Xie et al.   \cite{xieEffectiveMethodGenerate}                                          & Network   attack data generation            & Network security   systems require a large amount of data to perform network analysis                                                                                 & WGAN                                                                                                               \\ \hline
			\multirow{8}{*}{\textbf{Internet of   Things}} & Alshinina and   Elleithy \cite{alshininaHighlyAccurateMachine}                          & Wireless   Sensor Networks                  & \begin{tabular}[c]{@{}c@{}}Improve the   security of middleware in WSNs\\ Reduce power headroom\end{tabular}                                                  & DCGAN                                                                                                              \\ \cline{2-5} 
			& Nabati et al.   \cite{nabatiUsingSyntheticData}                                        & Indoor Localization                         & \begin{tabular}[c]{@{}c@{}}Shortcoming   of real data in WiFi fingerprinting localization methods\\Cost and time   consumption of collecting data\end{tabular} & Vanilla GAN                                                                                                        \\ \cline{2-5} 
			& Li et al.   \cite{liWaveletTransformDCGAN}                                              & Indoor Localization                         & \begin{tabular}[c]{@{}c@{}}Lack of diversity in the gathered CSI data\\Speed of convergence in the training phase and accuracy\end{tabular}                    & DCGAN                                                                                                        \\ \cline{2-5} 
			& Mohammadi et   al. \cite{mohammadiPathPlanningSupport}                                  & Path Planning                               & Generating   paths in wayfinding applications for disabled people                                                                                                     & Vanilla GAN   with a Classifier                                                                                    \\ \cline{2-5} 
			& Fard Moshiri   et al. \cite{moshiriUsingGANEnhance}                                     & Human   Activity Recognition                & Costs and   time consumption of collecting CSI data for HAR                                                                                                           & Vanilla GAN                                                                                                        \\ \cline{2-5} 
			& Xiao et al.   \cite{xiaoCsiGANRobustChannel}                                            & Human   Activity Recognition                & Low accuracy   of general approaches for left-out users                                                                                                               & Vanilla GAN and CycleGAN                                                                                                             \\ \hline
			\multirow{12}{*}{\textbf{Physical   Layer}}     & O'Shea et al.   \cite{osheaPhysicalLayerCommunications}                                & Approximation   of Channel Response         & Complexity of   modeling wireless channels                                                                                                                            & Vanilla GAN                                                                                                            \\ \cline{2-5} 
			& O'Shea et al.   \cite{osheaApproximatingVoidLearning}                                   & Approximation   of Channel Response         & Extension of   \cite{osheaPhysicalLayerCommunications}, learning the PDF of channels                                                                                                                     & Vanilla GAN                                                                                                  \\ \cline{2-5} 
			& Ye et al.   \cite{yeChannelAgnosticEndtoEnd}                                            & End-to-End   Communication Systems Model    & \begin{tabular}[c]{@{}c@{}}End-to-end   learning of a system without prior information\\Can be   applied to more realistic channels\end{tabular}               & CGAN                                                                                                               \\ \cline{2-5} 
			& Smith and   Downey \cite{smithCommunicationChannelDensity}                              & Channel   Density Estimation                & Learning   non-linearities, memory effects and non-Gaussian statistics in channels                                                                                    & BicycleGAN                                                                                                         \\ \cline{2-5} 
			& Yang et al. \cite{yangGenerativeAdversarialNetworkBasedWirelessChannel}                 & Wireless   Channel Modeling                 & Difficulty   and low accuracy of traditional channel modeling methods                                                                                                 & Vanilla GAN                                                                                                        \\ \cline{2-5} 
			& Zhao et al. \cite{zhaoClassificationSmallUAVs}                                          & UAV   Classification                        & Previous   methods to classify small UAVs required a large amount of samples for feature   extraction.                                                                & AC-WGAN                                                                                                            \\ \cline{2-5}
			& Davaslioglu   and Sagduyu \cite{davasliogluGenerativeAdversarialLearning}              & Spectrum   Sensing                          & \begin{tabular}[c]{@{}c@{}}Dynamicity of   wireless channels \\Shortcoming   of data for training a model for all environments\end{tabular}                    & CGAN                                                                                                               \\ \cline{2-5} 
			& Li et al.   \cite{liRadioClassifyGenerative}                                            & Modulation   Recognition                    & Increasing   the robustness of previous automatic modulation recognition frameworks                                                                                   & Modified DCGAN                                                                                                              \\ \cline{2-5} 
			& Shi et al.   \cite{shiGenerativeAdversarialNetwork}                                    & Wireless   Signal Spoofing                  & Generate   spoofing signals that are indistinguishable from intended signals                                                                                          & Vanilla GAN                                                                                                                \\ \cline{2-5} 
			& Erpek et al.   \cite{erpekDeepLearningLaunching}                                        & Wireless   Jamming                          & Dependency of   jamming on the amount of collected training data                                                                                                      & CGAN                                                                                                               \\ \cline{2-5}
			& Liao et al.    \cite{liaoGenerativeAdversarialNetwork}                                  & Covert Communication Systems                & Modeling the adversarial  game in covert communication with GANs and utilizing them for power allocation                                                                           & Vanilla GAN                                                                                                            \\ \hline
			\multirow{20}{*}{\textbf{Cybersecurity}}      & Usama et al.   \cite{usamaGenerativeAdversarialNetworks}                                & Intrusion   detection systems               & Vulnerability   of machine learning models to adversarial perturbations                                                                                                             & Vanilla GAN                                                                                                        \\ \cline{2-5} 
			& Lin et al.   \cite{linIDSGANGenerativeAdversarial}                                      & Intrusion   detection systems               & Increasing   robustness of intrusion detection systems                                                                                                                & WGAN                                                                                                             \\ \cline{2-5} 
			& Seo   et al. \cite{seoGIDSGANBased}                                                     & Intrusion   detection systems in vehicles   & Lack of   security features in CAN bus, reducing the false-positive error rate in   vehicle IDS                                                                       & Vanilla-GAN                                                                                                               \\ \cline{2-5} 
			& Salem et al. \cite{salemAnomalyGenerationUsing}                                        & Host-based   intrusion detection system     & \begin{tabular}[c]{@{}c@{}}Data imbalance and anomalies in host-based intrusion data sources,\\Increasing robustness of HIDS\end{tabular}                             & Cycle-GAN                                                                                                          \\ \cline{2-5} 
			& Amin et al.   \cite{aminAndroidMalwareDetection}                                        & Malware   detection in Android              & Growth of   Android operating system malware, limitations of learning-based malware   diagnosis techniques                                                            & LSTM-GAN                                                                                                           \\ \cline{2-5} 
			& Kim et al.   \cite{kimMalwareDetectionUsing}                                           & Malware   detection                         & Classification   and detection of zero-day attacks, detecting malware with a small amount of   data                                                                   & tGAN                                                                                                               \\ \cline{2-5} 
			& Roy et al.   \cite{royDetectionRogueRF}                                                 & Rogue RF   transmitter detection            & Limitations   of classical machine learning methods in detection of RF malicious activity                                                                             & Vanilla GAN                                                                                                        \\ \cline{2-5} 
			& Shin et al.   \cite{shinAndroidGANDefendingAndroid}                                     & Android pattern lock system            & Increasing the security of android pattern lock system by using the discriminator for anomaly detection                                                                    & Vanilla GAN                                                                                                        \\ \cline{2-5} 
			& Rigaki and   Garcia \cite{rigakiBringingGANKnifeFight}                                  & Malware   improvement                       & Avoiding   intrusion detection systems by modifying malicious traffic                                                                                                 & LSTM-GAN                                                                                                           \\ \cline{2-5} 
			& \begin{tabular}[c]{@{}c@{}}Hu and Tan   \cite{huGeneratingAdversarialMalware}\\Kawai et al.   \cite{kawaiImprovedMalGANAvoiding}\end{tabular} & Generating   adversarial malware examples   & \begin{tabular}[c]{@{}c@{}}Bypassing   black-box machine learning-based detection methods,\\Decreasing the detection  rate of malware\end{tabular} & Vanilla GAN                                                                                                            \\ \cline{2-5} 
			& Shi et al.   \cite{shiGenerativeAdversarialNetworks}                                    & Black-box API   attacks                     & Limited   amount  of training data due to limited access to the objective API                                                                                          & CGAN                                                                                                               \\ \cline{2-5} 
			& Chen et al.   \cite{chenCreditCardFraud}                                                & Credit card   fraud detection               & Highly skewed   datasets with little fraudulent  data                                                                                                                          & SAE and GAN                                                                                                        \\ \cline{2-5} 
			& Sethia et al.   \cite{sethiaDataAugmentationUsing}                                      & Credit card   fraud detection               & Highly class-imbalanced data                                                                                                                                        & \begin{tabular}[c]{@{}c@{}}Vanilla GAN,\\LSGAN,\\WGAN,\\MAGAN\\RWGAN\end{tabular}                                  \\ \cline{2-5} 
			& Wang et al.   \cite{wangFraudulentDataSimulation}                                       & Credit card   fraud detection               & Highly class-imbalanced data                                                                                                                                        & Vanilla GAN                                                                                                              \\ \cline{2-5} 
			& Hital et al.   \cite{hitajPassGANDeepLearning}                                          & Password   guessing                         & Difficulties   of implementing classic password guessing tools                                                                                                        & WGAN-GP                                                                                                            \\ \hline
		\end{tabular}%
	}
\end{table*}

\begin{table*}[h!]
	\centering
	\renewcommand{\arraystretch}{1.3}
	\caption{Parameters of the five different GANs evaluated in the experiment.}
	\label{tab:4}
	\resizebox{\textwidth}{!}{%
		\begin{tabular}{|c|c|c|c|c|c|}
			\hline
			& \textbf{Vanilla GAN} & \textbf{CGAN}       & \textbf{BIGAN}      & \textbf{LSGAN}     & \textbf{WGAN}    \\ \hline
			\textbf{Loss}             & Binary Crossentropy  & Binary Crossentropy & Binary Crossentropy & Mean Squared Error & Wasserstein Loss \\ \hline
			\textbf{Optimizer}        & Adam                 & Adam                & Adam                & Adam               & RMSprop          \\ \hline
			\textbf{Learning Rate}    & 0.0002               & 0.0002              & 0.0002              & 0.0002             & 0.00005          \\ \hline
			\textbf{Latent Dimension} & \multicolumn{5}{c|}{100}                                                                                 \\ \hline
			\textbf{Batch Size}       & \multicolumn{5}{c|}{64}                                                                                  \\ \hline
			\textbf{Epochs}           & \multicolumn{5}{c|}{5000}                                                                                \\ \hline
		\end{tabular}%
	}
\end{table*}

\section{Evaluation Framework}
\label{sec:4}
Thus far, we have reviewed the extant literature on GANs and their computer and communication networks application. In this section, we propose a suite of methods for measuring GAN performance with respect to our case study applications. We further train five state-of-the-art GAN models on four network-related datasets of different types and shapes and use the proposed methods, amongst other visualization tools, to evaluate model performance.
\subsection{Evaluation Metrics}
Evaluating and comparing the performance of GANs and other generative methods has always been a challenging task for researchers. Since GANs were mainly introduced for image data, the simplest and most straightforward evaluation method is a visual examination by humans, which is highly biased and subjective (different human judges are typically asked to look at pictures and vouch for the quality of generated images). However, this is inapplicable to data that does not readily fit human sensory categories. Other qualitative and quantitative assessment methods are available, however; Borji \cite{borjiProsConsGAN} presents 24 quantitative and 4 qualitative methods for evaluating and comparing GANs. The majority of these methods, such as Inception Score \cite{salimansImprovedTechniquesTraining}, Mode Score \cite{cheModeRegularizedGenerative} and Fréchet Inception Distance \cite{heuselGANsTrainedTwo} are only applicable to image data. From \cite{borjiProsConsGAN} and \cite{xuEmpiricalStudyEvaluation} we can conclude that out of these 28 proposed measures, only a small subset, including Average Log-likelihood \cite{goodfellowGenerativeAdversarialNets, theisNoteEvaluationGenerative}, Wasserstein Distance \cite{arjovskyWassersteinGAN} and Maximum Mean Discrepancy (MMD) \cite{grettonKernelTwosampleTest} are appropriate for non-image GAN data.
\par The task of evaluating generative methods is equivalent to measuring the dissimilarity between ${p_r}$ and ${p_g}$ (respectively, the probability distributions of samples drawn from real and generated data). In the case in which both of these distributions are known {\em a priori}, Kullback-Leibler Divergence (or Log-likelihood) and Jensen-Shannon divergence may be used, which are respectively defined as follows:
\begin{equation}
\label{eq:17}
{D_{KL}}({p_r}||{p_g}) \buildrel \Delta \over = \sum\limits_{x \in X} {{p_r}(x)\log \left( {\frac{{{p_r}\left( x \right)}}{{{p_g}(x)}}} \right)},
\end{equation}
\begin{equation}
\label{eq:18}
JSD({p_r}||{p_g}) = \frac{1}{2}{D_{KL}}({p_r}||M) + \frac{1}{2}{D_{KL}}({p_g}||M),
\end{equation}
where $X$ is the probability space and $M = \frac{1}{2}({p_r} + {p_g})$.
\par However, we generally do not have information about the distributions, but rather only access to finite samples drawn from them. Thus, we must estimate the destiny function of these probabilities. Kernel Destiny Estimation (KDE) is perhaps the most commonly used method for this task. For a probability kernel $K$ (such as a Gaussian) with bandwidth $h$ and independent and identically distributed (i.i.d) samples $\{ {x_1},{x_2},...,{x_n}\} $, the kernel destiny estimator is:
\begin{equation}
\label{eq:19}
\hat p\left( x \right) = \frac{1}{n}\sum\limits_{i = 1}^n {K\left( {\frac{{x - {x_i}}}{h}} \right)}.
\end{equation}
\par Although this measure is simple to compute; it suffers from a few drawbacks. Firstly, even for a large number of samples, KDE fails to approximate the model's true log-likelihood when the data dimensionality is high. Secondly, it may be shown that log-likelihood is uninformative regarding the quality of generated samples \cite{theisNoteEvaluationGenerative}. Consequently, a model that produces excellent samples may nonetheless have a poor log-likelihood. For these reasons, we will not use average log-likelihood as an evaluation metric for comparing different GAN models. 
\par The Wasserstein critic is an approximation version of the Wasserstein distance (also called Earth Mover Distance (EMD)) between 2 data distributions, ${P_r}$ and ${P_g}$ and is given in equation \cite{arjovskyWassersteinGAN, borjiProsConsGAN} below:
\begin{equation}
\label{eq:20}
W({p_r},{p_g}) \propto \mathop {\max }\limits_F {E_{x\sim{p_r}}}[F(\mathbf{x})] - {E_{x\sim{p_g}}}[F(x)],
\end{equation}
where $F$ is the Lipchitz function. Practically, the $F$ is an MLP with clipped weights. In realistic scenarios, the expectation is not taken; rather, the average over the Lipchitz function is utilized:
\begin{equation}
\label{eq:21}
\hat W({x_r},{x_g}) = \frac{1}{N}\sum\limits_{i = 1}^N {\hat F({x_r}[i])}  - \frac{1}{N}\sum\limits_{i = 1}^N {\hat F({x_g}[i])}.
\end{equation}
\par MMD is a measurement for comparing the dissimilarity between samples drawn from a pair of probability distributions and can be defined as a particular function space that witnesses the difference between the two distributions. Hence, a lower MMD implies that the two distributions are closer to each other. Thus if ${P_r}$ and ${P_g}$ are two probability distributions, the kernel MMD (squared MMD) between these two distributions will be \cite{grettonKernelTwosampleTest}:
\begin{equation}
\label{eq:22}
\begin{split}
MM{D^2}({p_r},{p_g}) =& {E_{x,{x^\prime }}}[k(x,{x^\prime })] - 2{E_{x,y}}[k(x,y)]\\ +& {E_{y,{y^\prime }}}[k(y,{y^\prime })],
\end{split}
\end{equation}
where $x$ and ${x}'$ are two independent random variables with distribution $p_r$, $y$ and ${y}'$ are two independent random variables with distribution $p_g$ and $k$ is a fixed characteristic kernel function.
\begin{figure*}[!t]
	\subfigure[]{\includegraphics[width=0.5\linewidth]{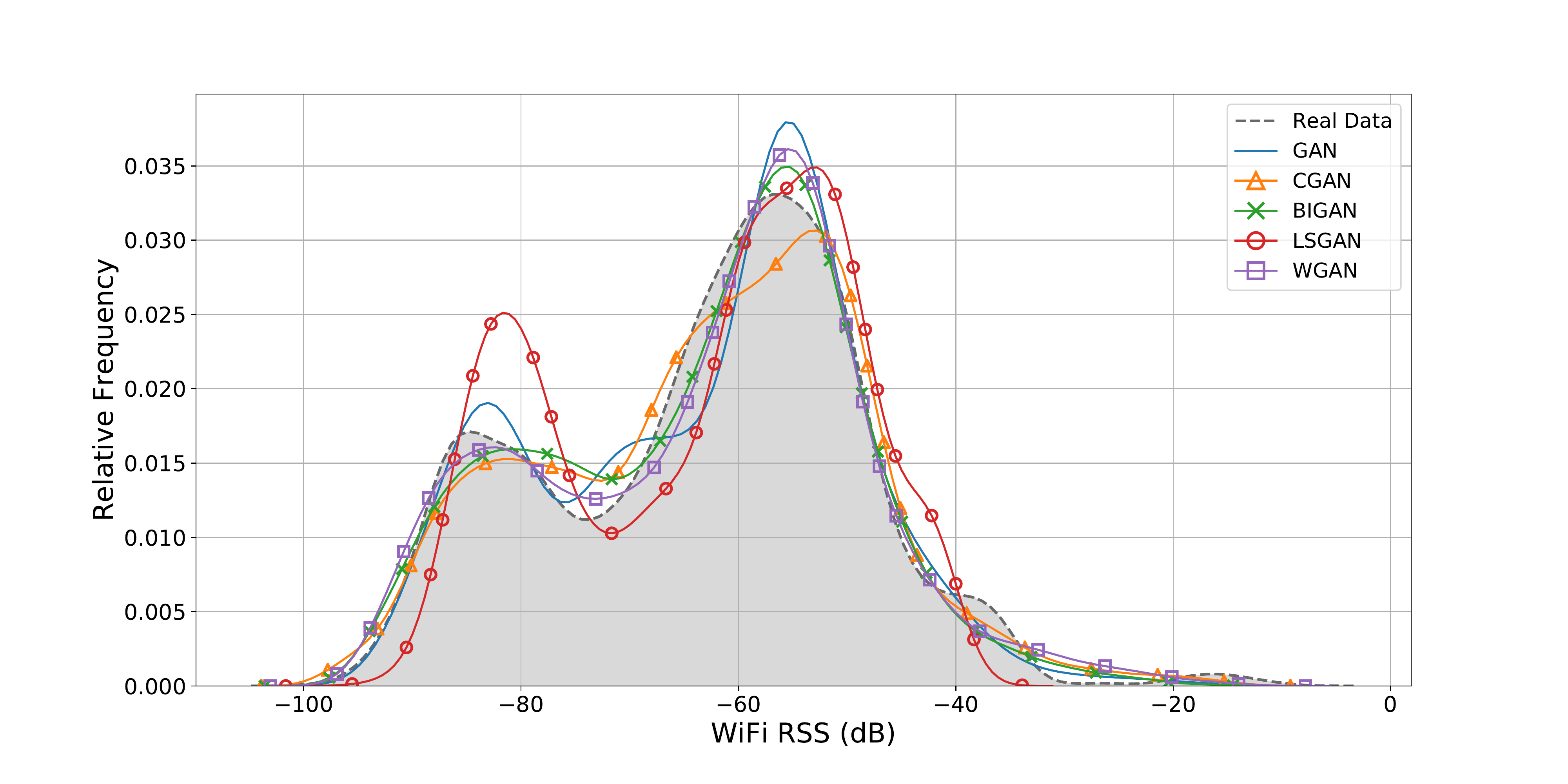}}
	\subfigure[]{\includegraphics[width=0.5\linewidth]{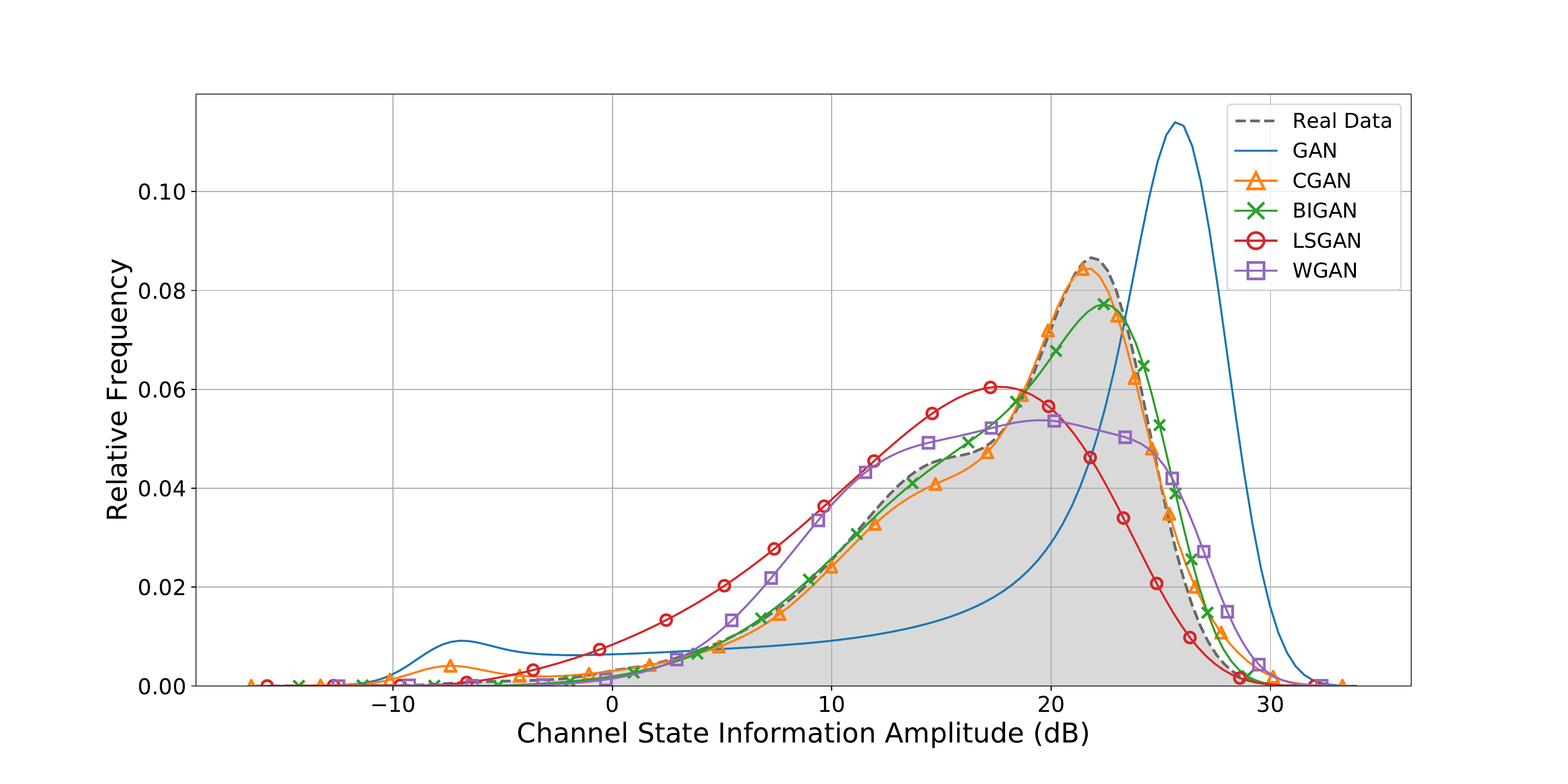}}
	\subfigure[]{\includegraphics[width=0.5\linewidth]{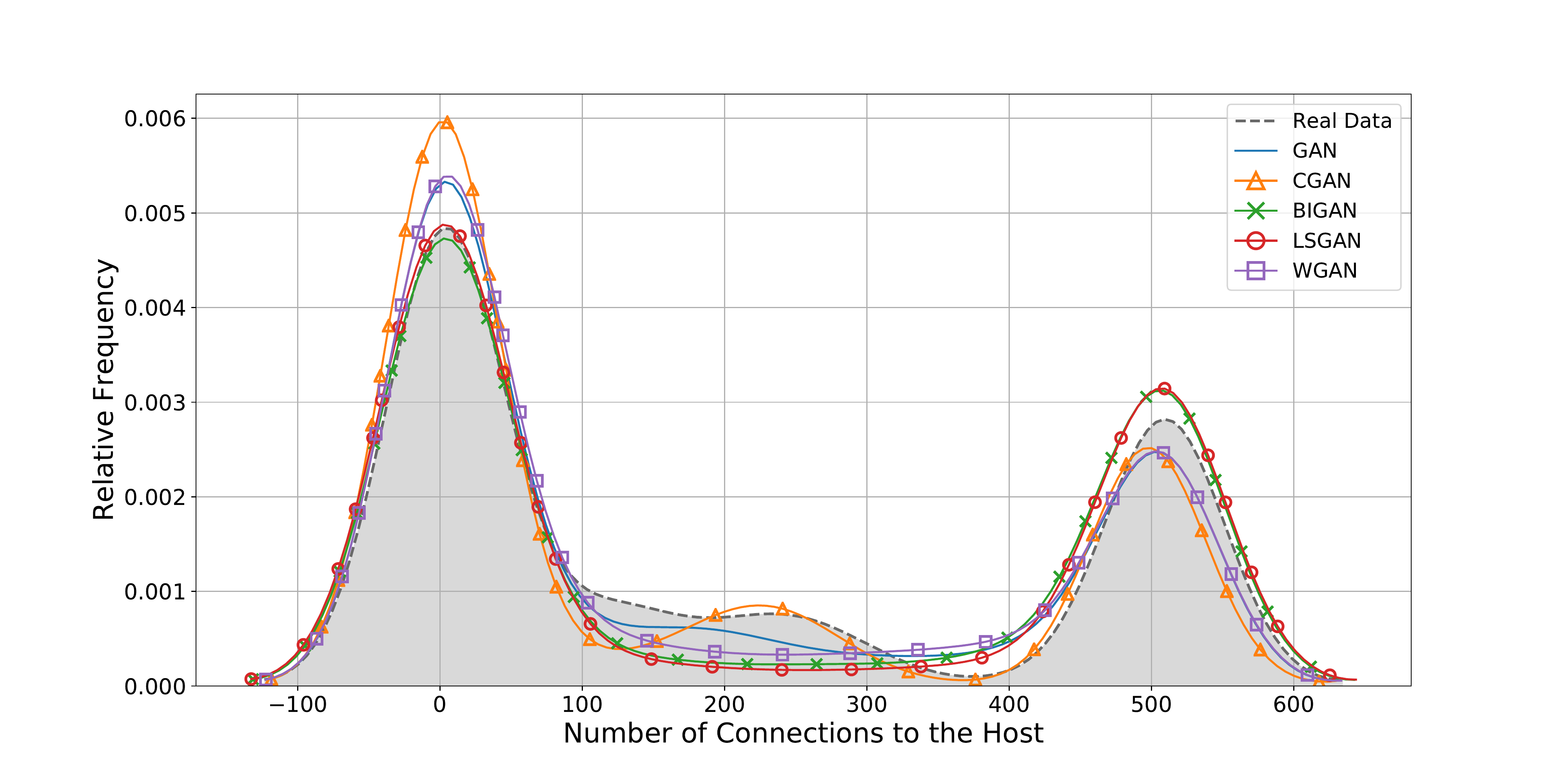}}
	\subfigure[]{\includegraphics[width=0.5\linewidth]{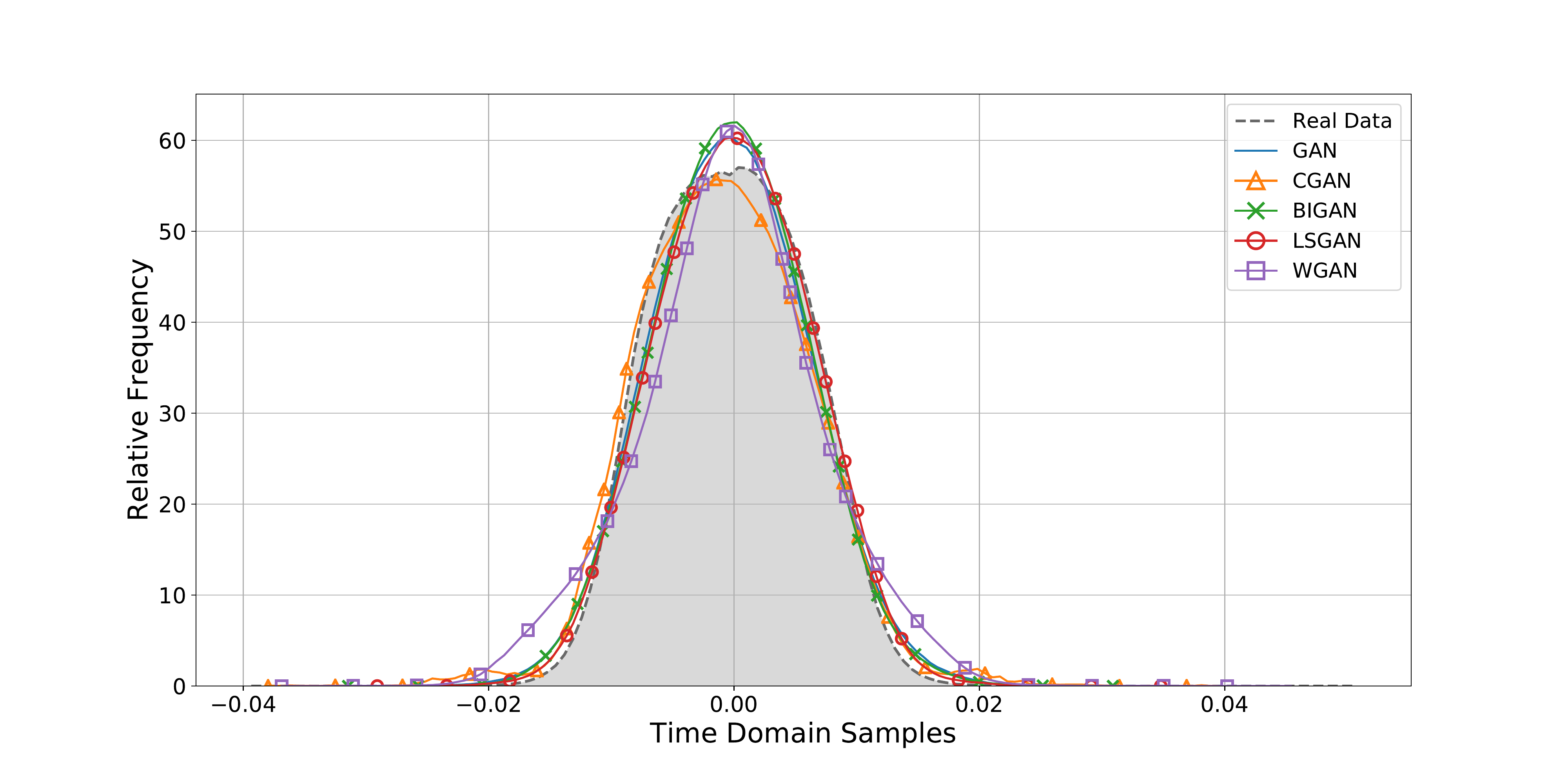}}
	\caption{Estimated probability distributions for a) RSS dataset, b) CSI dataset, c) KDD99 dataset, column ``count'', and d) RADIOML datasets. The shaded line is the estimated distribution of the real data, while the dashed lines are those of the synthetic data.}
	\label{fig:7}
\end{figure*}
\subsection{Experimental Methodology and Results}
\par In order to properly evaluate the performance of the various GANs within different network-related applications, we select reference datasets of varied types and characteristics. The datasets are: WiFi RSSI \cite{rohraUserLocalizationIndoor}, CSI \cite{yousefiSurveyBehaviorRecognition}, network traffic (KDD99) \cite{UCIKDDArchive} and digital modulations (Deepsig RadioML 2016.10A) \cite{incRFDatasetsMachine}. Similarly, five different types of GANs are selected for evaluation, including Vanilla GAN, CGAN, BIGAN, LSGAN, and WGAN. The reason for this choice is that they are the most used models in the literature. Furthermore, while being easy to train with non-image data, each of these models adds a fundamentally different aspect to the original Vanilla GAN model. For instance, CGAN makes use of data labels, while BIGAN is able to map the real data distribution to the latent space. Experiments are carried out in Keras, accelerated by a Geforce RTX 2060 GPU. To keep consistency between implementations of different GANs and to mitigate possible errors that may occur in the implementation phase, we use Keras-GAN\footnote{https://github.com/eriklindernoren/Keras-GAN} as our base framework. The respective GAN model architecture parameters (shown in Table \ref{tab:4}) are not adjusted during the experimental process, and no hyperparameter tuning is done (as altering learning parameters could bias the evaluation process).
\par We train each of the five GANs on each of the datasets separately and use noise drawn from the {\em same} latent space to generate samples. All of the datasets, with the exception of KDD99, contain data of the same type and magnitude. For KDD99, since GANs cannot generate discrete data, we select only 18 of the continuous non-zero features and use these to train the models. To make the experiment reproducible, we use the same random seed throughout the experiment for all datasets and models. However, since the GANs are trained over a sufficient amount of epochs, and the datasets are sufficiently large, randomness would have minimal effect on the results. Comparing performance between models is conducted via measurement of the two metrics introduced above; MMD and EMD. The evaluation process results for all of the datasets are as depicted in Table \ref{tab:5}.
\par To further visualize these experimental outcomes, we use KDE to estimate the distribution of real and generated data. The estimated distributions of all four datasets are depicted in Fig.\ref{fig:7}. It should be noted that the selected features of the KDD99 datasets are of different orders, as some are of packet length, being in the order of 10000s, while others are of the order of 100s and even 10s. For this reason, in order to make the distribution plot more readable without loss of generality, we depict only the distribution for one of the columns, ``count,'' which is the number of simultaneous connections to the host. 
\par
\begin{figure*}[!t]
	\subfigure[]{\includegraphics[width=0.5\linewidth]{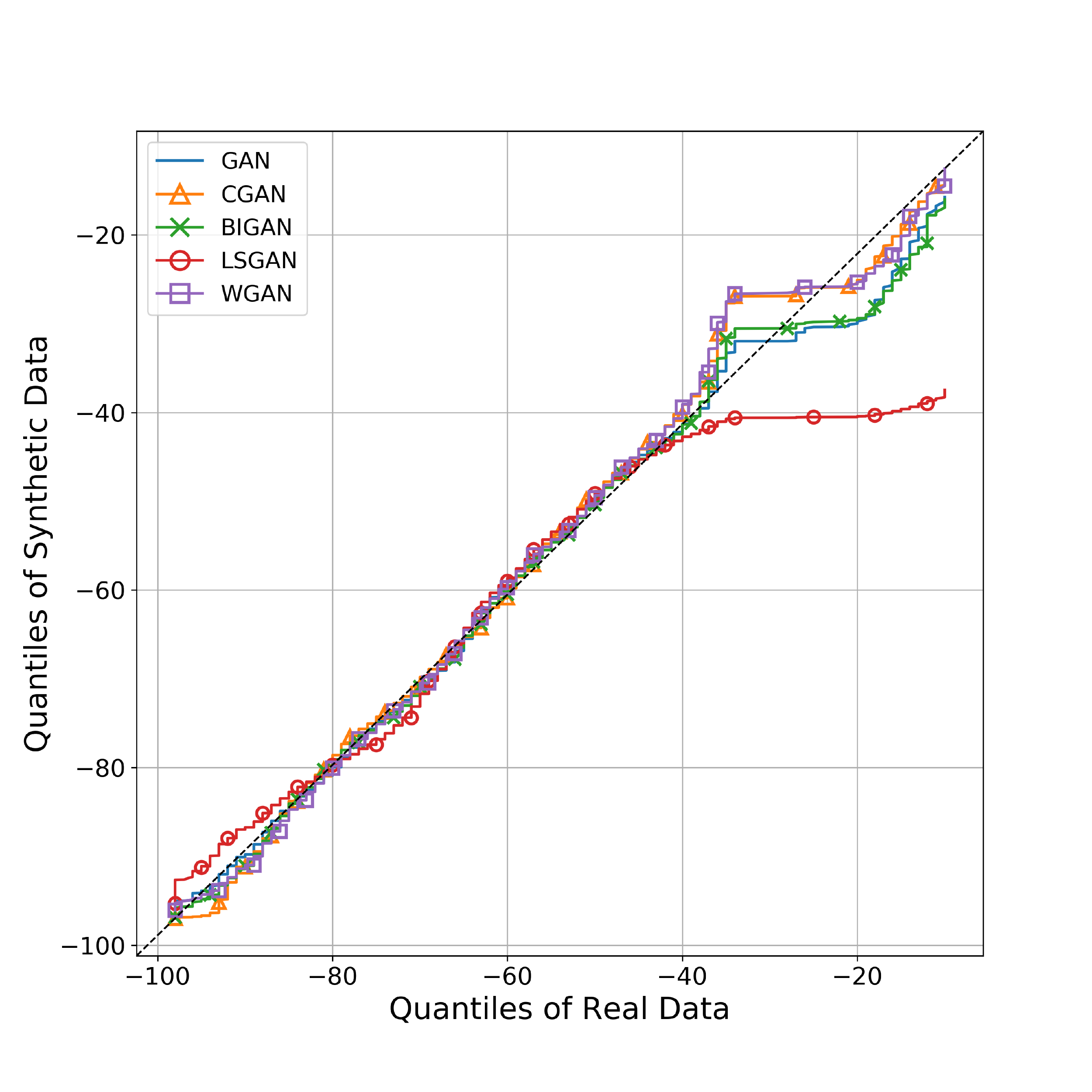}}
	\subfigure[]{\includegraphics[width=0.5\linewidth]{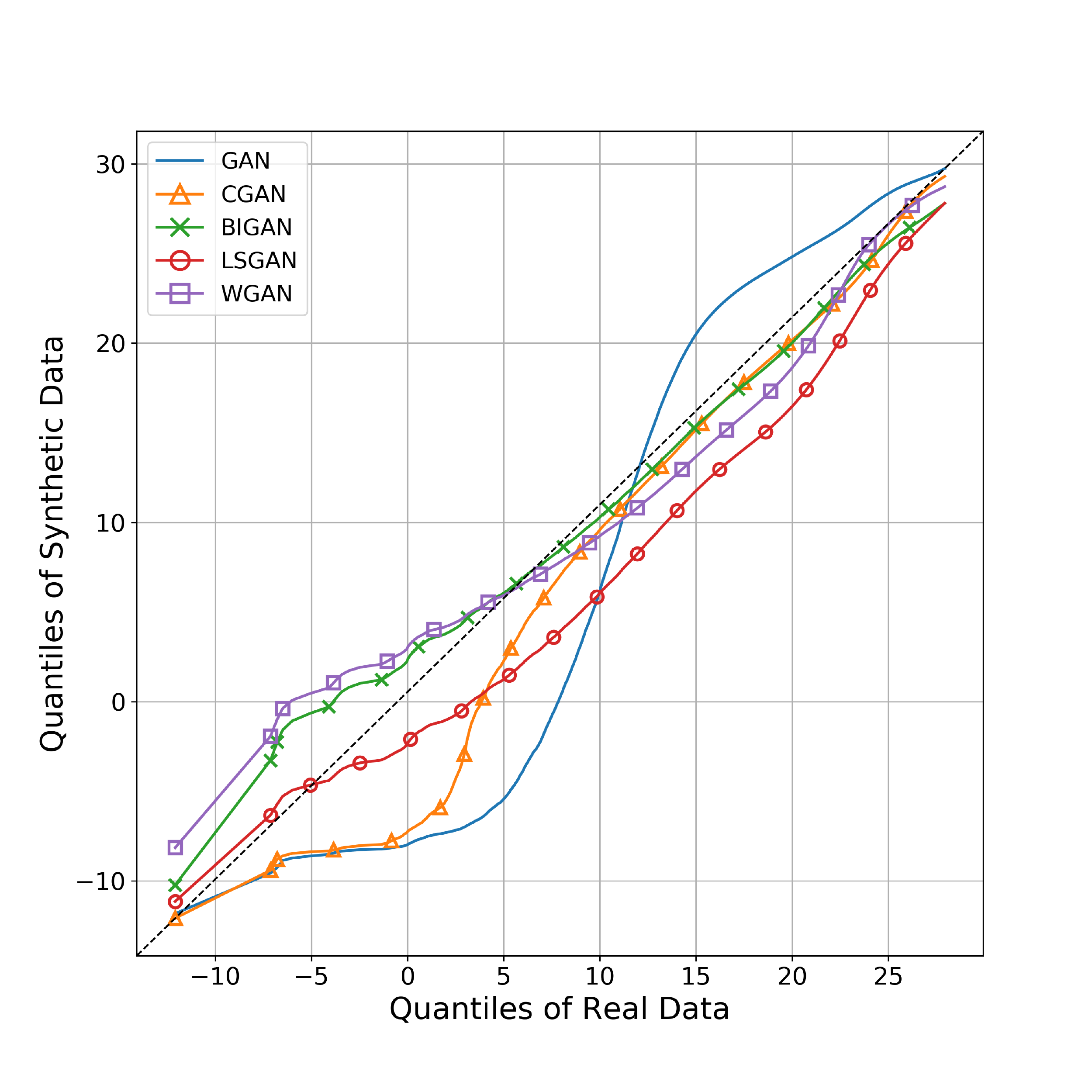}}
	\subfigure[]{\includegraphics[width=0.5\linewidth]{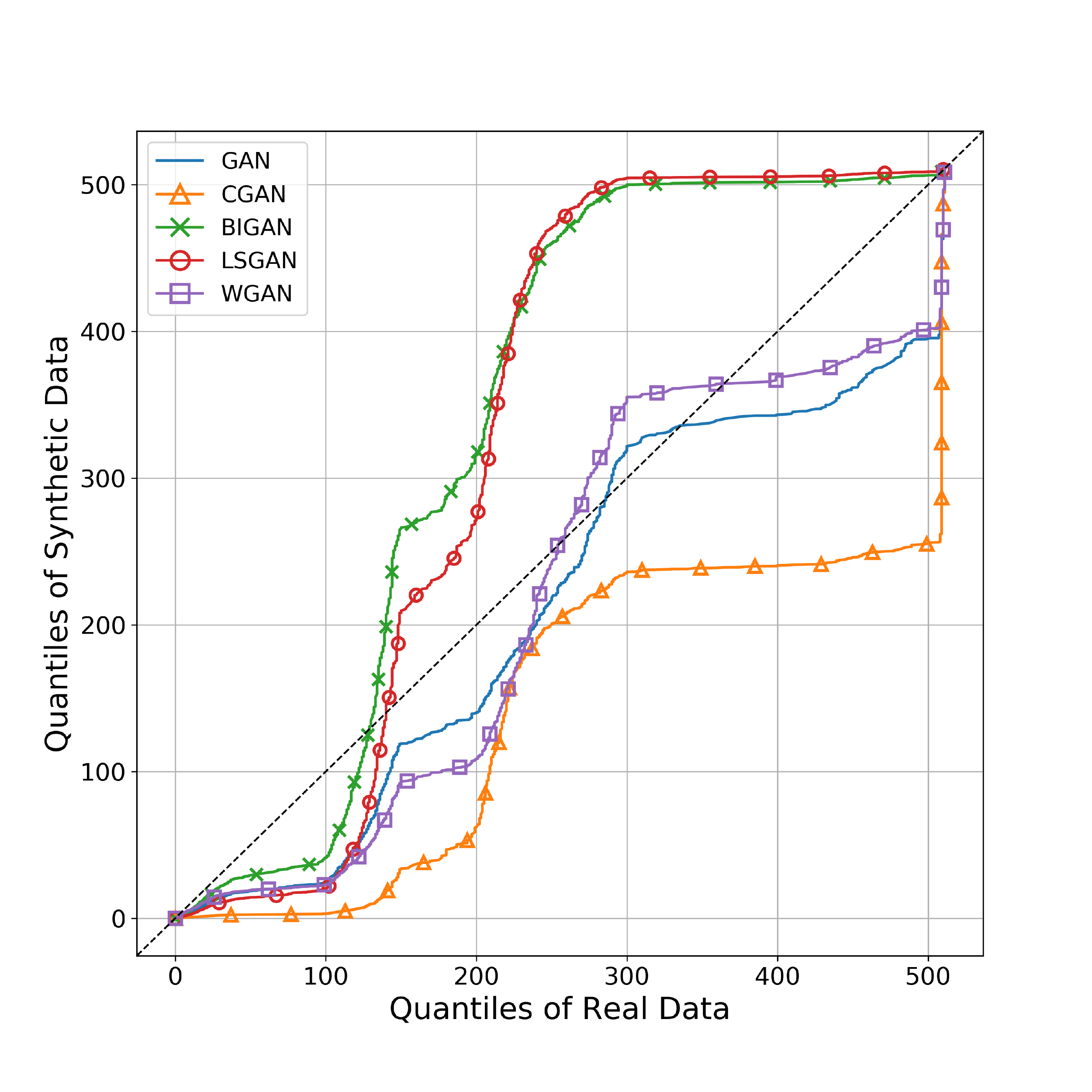}}
	\subfigure[]{\includegraphics[width=0.5\linewidth]{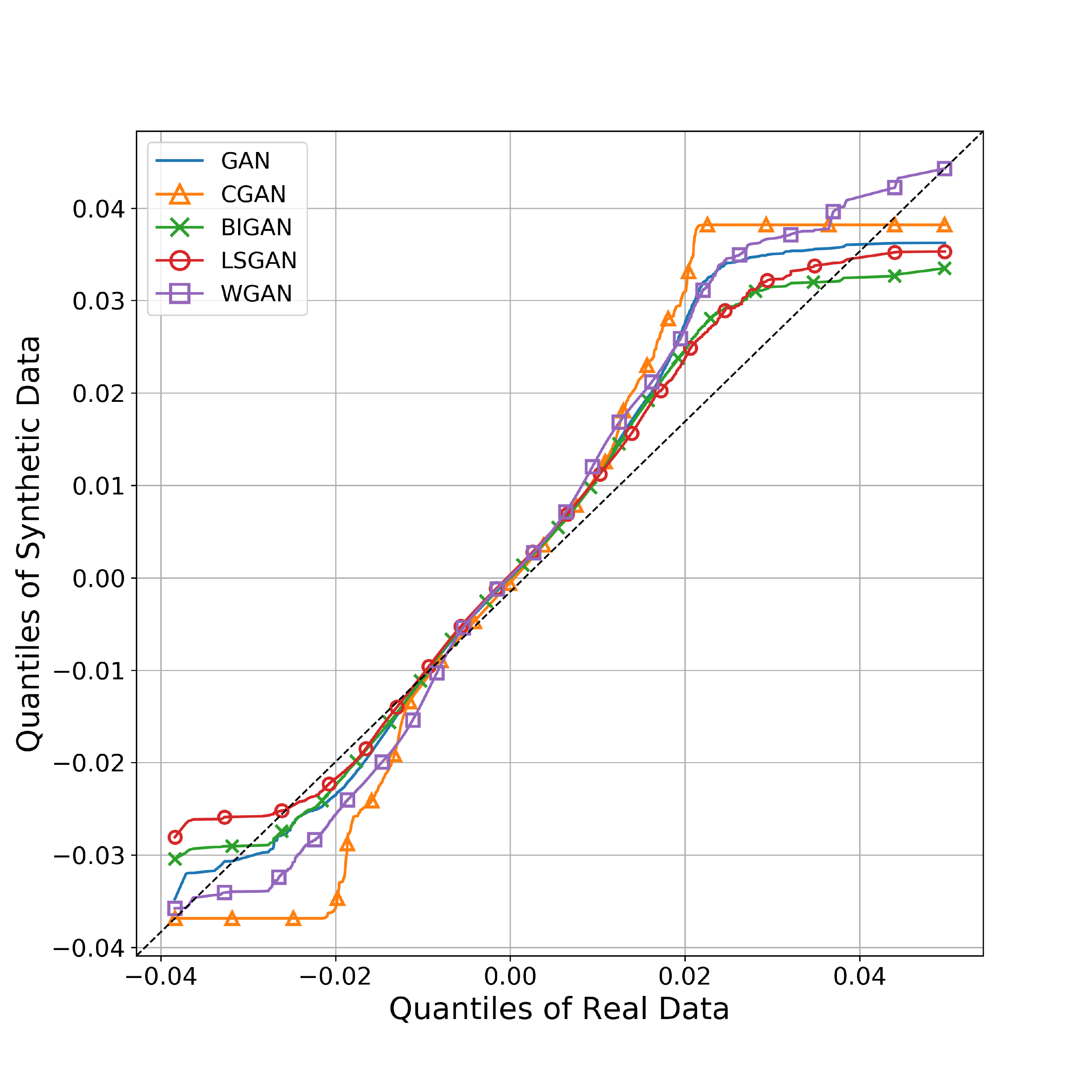}}
	\caption{Q-Q plot for a) RSS dataset, b) CSI dataset, c) KDD99 dataset, column ``count'', and d) RADIOML datasets.}
	\label{fig:8}
\end{figure*}

 The final measure we consider is the quantile-quantile (Q-Q) plot, a graphical nonparametric method to compare the shape of two probability distributions. It consists in a scatter plot of the quantiles of one dataset against the quantiles of the other. If the points are close to the 45-degree reference line, we may conclude that the datasets are sampled from similarly-shaped distributions (although they may have different underlying parameters). Contrarily, if the points are far from the reference line, we may conclude that the distributions differ significantly. While the Q-Q plot can be considered a visual guide for comparing the two datasets' similarities, it should not be taken as reliable proof of similarity in itself. However, its ability to relatively characterize statistical properties such as central tendency, dispersion, and skewness makes it an extremely useful tool. The Q-Q plots of the datasets, as depicted in Fig.\ref{fig:8}, visually demonstrate that the estimated distributions shown in Fig.\ref{fig:7}, are indeed accurate.
\begin{table}[h]
	\centering
	\renewcommand{\arraystretch}{1.2}
	\caption{Evaluation metrics for five different GANs on the four reference datasets. The used metrics are MMD and EMD, where a lower value means the generated data are closer to the original data.}
	\label{tab:5}
	\resizebox{\columnwidth}{!}{%
		\begin{tabular}{|c|c|c|c|c|}
			\hline
			\textbf{Dataset}                  & \textbf{Data Shape}            & \textbf{Model} & \textbf{MMD} & \textbf{EMD} \\ \hline
			\multirow{5}{*}{WiFi RSSI \cite{rohraUserLocalizationIndoor}}  & \multirow{5}{*}{(2000,7)}      & Vanilla GAN    & 0.077095     & 7.699772     \\ \cline{3-5} 
			&                                & CGAN           & 0.064150     & 9.186863     \\ \cline{3-5} 
			&                                & BIGAN          & 0.067965     & 9.912520     \\ \cline{3-5} 
			&                                & LSGAN          & 0.132464     & 10.56940     \\ \cline{3-5} 
			&                                & WGAN           & 0.042694     & 8.440702     \\ \hline
			\multirow{5}{*}{WiFi CSI \cite{yousefiSurveyBehaviorRecognition}} & \multirow{5}{*}{(2100,500,90)} & Vanilla GAN    & 0.602905     & 1844.109     \\ \cline{3-5} 
			&                                & CGAN           & 0.129852     & 810.8793     \\ \cline{3-5} 
			&                                & BIGAN          & 0.108452     & 745.0461     \\ \cline{3-5} 
			&                                & LSGAN          & 0.476095     & 1589.919     \\ \cline{3-5} 
			&                                & WGAN           & 0.158754     & 778.3315     \\ \hline
			\multirow{5}{*}{KDD99 \cite{UCIKDDArchive}}      & \multirow{5}{*}{(6000,18,1)}   & Vanilla GAN    & 0.036804     & 697.4859     \\ \cline{3-5} 
			&                                & CGAN           & 0.178606     & 8423.204     \\ \cline{3-5} 
			&                                & BIGAN          & 0.061430     & 994.2017     \\ \cline{3-5} 
			&                                & LSGAN          & 0.147254     & 3376.442     \\ \cline{3-5} 
			&                                & WGAN           & 0.029684     & 749.5329     \\ \hline
			\multirow{5}{*}{RadioML \cite{incRFDatasetsMachine}}    & \multirow{5}{*}{(11000,2,128)} & Vanilla GAN    & 0.084636     & 0.108324     \\ \cline{3-5} 
			&                                & CGAN           & 0.152164     & 0.110944     \\ \cline{3-5} 
			&                                & BIGAN          & 0.107003     & 0.107230     \\ \cline{3-5} 
			&                                & LSGAN          & 0.097895     & 0.107172     \\ \cline{3-5} 
			&                                & WGAN           & 0.085040     & 0.113345     \\ \hline
		\end{tabular}
	}
\end{table}

\subsection{Results}
Inspection of the experimental results suggests that none of the tested GANs has a decisive advantage over the others. Performance heavily depends on properties such as the dimensionality, magnitude, morphology, and type of data used (as might be expected {\em a priori} given the `no free lunch' theorem). In other words, no variant of GAN can achieve superior performance for every type of data and in every application. Hence, the suitable way to find the proper model is through trial-and-error. For instance, Table \ref{tab:5} suggests that LSGAN and Vanilla GAN are not practical for high dimensional datasets, such as CSI, as they yielded a relatively large MMD and EMD. On the other hand, they indicate acceptable performance on other types of data.
\par Nevertheless, we assume that by changing the architecture of models (in particular, the number and depth of layers and the hyperparameter tuning), this situation could vary significantly. A further significant point to note from the results is that the tested GAN models cannot capture very fast fluctuations in the data distribution, presumably due to intrinsic model bias. It might be that the models are not sufficiently deep and that a deeper model with more layers and neurons would be able to capture these; however such architectural complexity has not as yet been applied practically in the network field.
\section{Discussion and Conclusion}
We have provided a survey of GANs that aims to be comprehensive regarding the main model variants set out in the general machine learning literature,
and exhaustive with respect to their application in the computer and communication networks domain. Despite being a relatively newly-proposed approach, GANs have been widely accepted in the machine learning community, with the quantity of research carried out in respect of them increasing at a significant pace. Therefore, we briefly introduced the concept of generative machine learning prior to comparing the structure of the principle GAN model variants against each other. Next, we divided the extant work in the literature into five main categories and reviewed various applications of the different models within these categories. Afterward, we introduced a selection of the quantitative evaluation metrics in use in the field and evaluated the performance of a range of different GAN models on a selection of datasets taken from the reviewed papers. Finally, we set out the major challenges and shortcomings of current GANs, and gave a preview of work to be done in the future. We strongly believe that GANs have many more potential applications within the field of computer and communication networks, and this discovery process has only just begun.
\subsection{Take-home lessons}
Alongside their extensive application in computer vision, GANs have a wide range of applications within networking. These vary from traffic classification, modulation recognition, self-organizing networks, spectrum sensing for intrusion detection systems, malware detection, and IoT. Generally, we benefit from the use of GANs where there are either data shortcomings, data imbalance, or else exceptional or adversarial circumstances for which an accurate discriminator is required, such as malware detection. When utilizing existing image GANs to generate non-image data, one of two broad approaches can be selected; we can convert the CNN models that are used in the majority of state-of-the-art GANs to generic multi-layer perceptions and train the network using normalized data, or else we can convert our data to images and use the existing GAN as is. An example of the former method is \cite{salemAnomalyGenerationUsing}. Whether one approach is superior to the other is likely to depend on the nature of the data.
\par As the training process of GANs does not require labels (except the case of conditional networks), GANs can be of enormous benefit in semi-supervised learning. For instance, one can train the GAN with the large portion of unlabeled data and then use the small portion of labeled data to train the discriminator for classification and regression; as done in \cite{liRadioClassifyGenerative}. There are hence many more applications in the field of networking that could potentially benefit from the discriminator network component of GANs.
\subsection{Future Work}
Beyond the above usages, there are many more network applications that could benefit from this generative approach. In particular, there are many areas in the fields of physical layer, wireless sensor networks, and mobile networks that could significantly benefit. At present, though, most of the available studies on GANs are for the purpose of image generation and translation. Even when they can be used for non-image data (see above), most of the existing evaluation metrics are developed for image data. This can make selecting the correct GAN a time-consuming process, as the only absolutely reliable way to make such a decision is to generate data from the different models and compare the associated classification accuracy of the destination model. Thus, introducing new theoretical and statistical metrics for comparing non-image data would have a potentially huge impact. 
\par A further point to note is that, although GANs were initially designed for image data, they can, by appropriate hyperparameter optimization of (or minor modification to) the underlying NNs, be made to generate practically any continuous multi-dimensional numerical data. However, as of yet, they are unable to convincingly generate discrete data in arbitrary environments, mainly because the generator network is not able to use back-propagation in this case (recently the boundary-seeking GAN (BSGAN) \cite{hjelmBoundarySeekingGenerativeAdversarial} has been proposed for this task, but the performance of this model has yet to be evaluated with respect to discrete data related to networking).
\par More generally, while the evolving variants of Vanilla GAN have improved on the basic architecture in terms of overall convergence reliability in providing solutions to  mode collapse and instability, GANs have not yet evolved to the extent of generating data that is uniformly indistinguishable both to humans and machines; this is still an open challenge. 
\par Finally, almost all of the reviewed work in this survey utilized GANs for supervised, unsupervised, and semi-supervised tasks; however, employing GANs in reinforcement learning is also a promising research direction. Recently, the amount of research addressing this has increased rapidly, demonstrating that GANs can be successfully combined with policy gradient \cite{yuSequenceGenerativeAdversarial}, imitation learning \cite{hoGenearitveAdversarialImitation}, and actor-critic methods \cite{pfauConnectingGenerativeAdversarial}. However, as of now, utilizing GANs with reinforcement learning algorithms in network-related tasks is something that remains to be done.

\bibliographystyle{unsrtnat}
\bibliography{refs}

\end{document}